\pdfoutput=1
\documentclass[bibyear]{aa}
\usepackage[colorlinks=true,     linkcolor=blue, citecolor=blue, filecolor=blue, urlcolor=blue]{hyperref}
\usepackage{natbib}
\usepackage{graphicx}
\usepackage{multirow}
\usepackage{makecell}
\usepackage{CJK}
\usepackage{txfonts}
\newcommand{\mjup}{~$\textit{M}_{\mathrm{J}}$}
\newcommand{\prim}{~PDS~70~}
\newcommand{\rin}{$R_{in}$}
\newcommand{\rout}{$R_{out}$}
\newcommand{\hrr}{$H(r)$}
\newcommand{\hr}{$H$}
\newcommand{\hra}{$H1$}
\newcommand{\hrb}{$H2$}
\newcommand{\hrc}{$H3$}
\newcommand{\plh}{~$\beta$~}
\newcommand{\plha}{~$\beta1$~}
\newcommand{\plhb}{~$\beta2$~}
\newcommand{\plhc}{~$\beta3$~}
\newcommand{\mdisk}{$M$}
\newcommand{\mdiska}{$M1$}
\newcommand{\mdiskb}{$M2$}
\newcommand{\mdiskc}{$M3$}
\newcommand{\plsig}{$\gamma$}
\newcommand{\plsiga}{~$\gamma1$~}
\newcommand{\plsigb}{~$\gamma2$~}
\newcommand{\plsigc}{~$\gamma3$~}
\newcommand{\submm}{sub-millimeter}
\newcommand{\chisq}{~$\chi^2$~}
\newcommand{\chisqR}{${\chi_R}^2$}
\newcommand{\mic}{~$\mu$m~}
\newcommand{\msun}{~$M_\odot$~}
\newcommand{\mearth}{~$M_\oplus$~}

\begin{document}

\title{PDS 70 Unveiled by Star-hopping: \\total intensity, polarimetry and mm-imaging modeled in concert}
\titlerunning{PDS 70 Unveiled by Star-hopping.}

\author{
    Wahhaj, Z. \inst{1},
    Benisty, M. \inst{2,3}, 
    Ginski, C. \inst{4},
    Swastik, C. \inst{5,1},
    Arora, S. \inst{6,1},
    van Holstein, R. G. \inst{1},
    De Rosa, R. J. \inst{1}, \\
    Yang, B. (\begin{CJK*}{UTF8}{gbsn}杨彬\end{CJK*}) \inst{7}, 
    Bae, J. \inst{8},
    Ren, B. (\begin{CJK*}{UTF8}{gbsn}任彬\end{CJK*})\inst{2}
}

   \institute{European Southern Observatory, Alonso de C\'ordova 3107, Vitacura Casilla 19001, Santiago, Chile 
        \email{zwahhaj@eso.org}
        \and
    Universit\'{e} C\^{o}te d'Azur, Observatoire de la C\^{o}te d'Azur, CNRS, Laboratoire Lagrange, Bd de l'Observatoire, CS 34229, 06304 Nice cedex 4, France
    \and
    Universit\'{e} Grenoble Alpes, CNRS, Institut de Plan\'{e}tologie et d'Astrophysique (IPAG), F-38000 Grenoble, France
        \and
        School of Natural Sciences, University of Galway, University Road, H91 TK33 Galway, Ireland 
        \and
        Indian Institute of Astrophysics, India,
        \and
        University of Potsdam, Germany,
        \and
        Instituto de Estudios Astrof\'{i}sicos, Facultad de Ingenier\'{i}a y Ciencias, Universidad Diego Portales.
        \and
        Department of Astronomy, University of Florida, Gainesville, FL 32611, USA
    }
   
   \authorrunning{Wahhaj, Benisty, Ginski et al.\ 2024} 
   
   \date{Submitted to A\&A on December 19, 2023. \\ Accepted on April 11, 2024}

 
  \abstract
   {Most ground-based planet search direct imaging campaigns use angular differential imaging, which distorts the signal from extended sources like protoplanetary disks. In the case PDS~70, a young system with two planets found within the cavity of a protoplanetary disk, obtaining a reliable image of both planets and disk is essential to understanding planet-disk interactions.}
   {Our goals are to reveal the true intensity of the planets and disk without self-subtraction effects for the first time, search for new giant planets beyond separations of 0.1$''$ and to study the morphology of the disk shaped by two massive planets.}
   {We present $YJHK$-band imaging, polarimetry, and spatially resolved spectroscopy of PDS 70 using near-simultaneous reference star differential imaging, also known as star-hopping. We created a radiative transfer model of the system to match the near-infrared imaging and polarimetric data, along with sub-millimeter imaging from ALMA. Furthermore, we  extracted the spectra of the planets and the disk and compared them.}
   {We find that the disk is quite flared with a scale height of $\sim$ 15\% at the outer edge of the disk at $\sim$ 90~au, similar to some disks in the literature. The gap inside of $\sim$50~au is estimated to have $\sim$1\% of the dust density of the outer disk. The Northeast outer disk arc seen in previous observations is likely the outer lip of the flared disk. Abundance ratios of grains estimated by the modeling indicate a shallow grain-size index $> -2.7$, instead of the canonical -3.5. There is both vertical and radial segregation of grains. Planet $c$ is well separated from the disk and has a spectrum similar to planet $b$, clearly redder than the disk spectra. Planet $c$ is possibly associated with the sudden flaring of the disk starting at $\sim$ 50~au. No new planets $>$ 5\mjup\ were found.} 
   {
   }


    \maketitle
%

\section{Introduction}

We have detected thousands of mature planetary systems (ages $>$ 1 Gyr)  with exo-neptunes, jupiters, ( and some near exo-earths) whose planetary properties have been analyzed to understand the end-points of the planet formation process \citep{Zhu:2021}. We have also detected hundreds of young systems (ages $<$ 0.1 Gyr) which form an image of the embryonic phases of the planet creation story. Recently, we have started to discover systems going through the planetary birth process \citep{2020ARA&A..58..483A}. Since this phase is short ($\lesssim$ 10~Myrs), such systems are rare \citep{Benisty:2023}. PDS 70 is a prime example of such a system.

The PDS 70 proto-planetary system is in the midst of forming at least two giant planets  \citep{keppler:2018, Haffert:2019} from a disk of gas and dust \citep{Dong:2012}. This system represents an invaluable trove of clues to understanding the evolutionary steps of the disk during the planet formation process \citep{2019ApJ...884L..41B}. The age of the PDS 70 system is estimated to be around 5 Myr, placing it beyond typical disk lifetimes.  See \citet{Muller:2018} \citep[also][]{Haffert:2019,2022ApJ...938..134S} for a discussion of the age of the system, and the related estimates of planet mass. It is a K7 pre-main sequence star with a mass of \(0.82\) solar masses \citep{Riaud:2006}, and [Fe/H]~$=$ -0.11$\pm$0.01~dex \citep{Swastik:2021}. The Gaia EDR3 motions of PDS 70 are consistent with it being a member of the Upper Centaurus Lupus star-forming region. The membership probability is as high as \(98.7\%\), as determined using Banyan \(\Sigma\) \citep{Gagne:2018}. The age of PDS 70 is significantly younger than the mean age of stars in the Upper Centaurus Lupus region (16 $\pm$ 2~Myr), but given that the instrinsic age spread is estimated to be $\sim$8~Myr, this is not surprising   \citep{Pecaut:2016}.

The system hosts two known planets, PDS 70 $b$ and PDS 70 $c$, both of which are still in their formation stages \citep{Benisty:2021,Mesa:2019}. Their masses are estimated to be less than 10\mjup \citep{Wang:2021,Mesa:2019}. The planets are located at approximately 22 and 34~au from the central star \citep{Muller:2018}. Atmospheric models applied to the planets suggest a temperature range of approximately 1000--1600~K and a surface gravity $\log(g) \leq 3.5$ dex \citep{Muller:2018}. \citet{Wang:2021} found that requiring dynamically stable orbits results in a $95\%$ upper limit on PDS 70 b's mass of 10\mjup. Their GRAVITY K-band spectra suggest dusty planetary atmospheres due to accreting dust. \citet{Mesa:2019} provide additional details, estimating PDS 70 $c$'s mass to be less than 5\mjup, with an effective temperature of around 900$K$ and a surface gravity, $\log(g)$, between $3.0$ and $3.5$ dex. Using ALMA data, \citet{Benisty:2021,Isella:2019} confirm the presence of circumplanetary disks (CPDs) around both planets, with emission around PDS 70 $c$ corresponding to a dust mass of approximately 0.031\mearth for 1\mic sized grains or 0.007\mearth for 1$mm$ sized grains. However, the CPDs are not easy to detect in polarized light \citep{vanHolstein:2021}. \citet{Haffert:2019} and \citet{Wagner:2018} both report strong H-alpha emission from the planets, indicating ongoing accretion. \citet{Wagner:2018} estimate the mass accretion rate for PDS 70 $b$ to be $10^{-8} M_J/\text{yr}$ to $10^{-7} M_J/\text{yr}$, suggesting that the planet has acquired approximately $90\%$ of its mass based on its current mass and accretion rate.

Both planets are thought to be in a 2:1 mean motion resonance, which has remained stable over millions of years \citep{2019ApJ...884L..41B,Haffert:2019}. PDS 70 $b$ has an eccentricity of 0.17$\pm$0.06, while PDS 70 $c$ has a near-circular orbit \citep{Wang:2021}.
PDS 70 $b$'s orbit is coplanar to the disk, with a period of 118 years \citep{Muller:2018}. 

The disk's gas and dust components have been extensively studied. ALMA observations with a spatial resolution of approximately 50 au have detected 16 transitions from 12 molecular species, including CO isotopologues, formaldehyde, small hydrocarbons, HCN, and HCO$^{+}$ isotopologues \citep{Facchini:2021}. ALMA observations have also revealed a radial dust gap at 0.42$"\pm$0.05 and peak densities at 0.75$"$ for the dust, 0.5$"$ for HCO$^{+}$ J=3-4, and 0.46$"$ for CO J=3-2 \citep{Facchini:2021,Long:2018}. The gas disk exhibits a shallower gap depth compared to the dust. Also, the gas spreads wider than the dust disk \citep{Facchini:2021,Long:2018}. The disk is also characterized by strong X-ray and UV radiation absorption near the star, shielding the outer disk \citep{2022ApJ...938..134S}. \citet{Cridland:2023} used a thermo-chemical code called DALI \citep{2012A&A...541A..91B, 2013A&A...559A..46B} to model the radial profiles of $^{12}$CO, C$^{18}$O, and C$_{2}$H. They found a Carbon-to-Oxygen (C/O) ratio greater than 1 in the outer disk, corroborating \citet{Facchini:2021}'s findings of a high C$_{2}$H/$^{12}$CO flux ratio. This suggests that the disk, and potentially the planets forming within it, could have a super-stellar C/O ratio in their atmospheres. Since the C/O ratio should change with distance from the star due to different condensation temperatures for the gases found, this could have implications for where the planets formed \citep{Cridland:2023}.

In this work, we present near-infrared (NIR) observations of the system obtained with the SPHERE instrument at the Very Large Telescope (VLT). These include $YJHK$-band total intensity imaging and spectroscopy, and $H$-band polarimetry. We use these along with ALMA observations from \citet{Benisty:2021} to create a consistent radiative transfer model of the disk. We then compare the astrometric and spectroscopic properties of the planets to the model disk properties to understand the interactions of the planets and the disk. The paper is organized as follows. In Section 2, we describe the observations and data reduction. Section 3 details the extraction of the planet photometry, astrometry, and spectra. We also extract the disk spectra. We use the astrometry to find the most likely orbital solutions for the planets. Section 4 describes the radiative transfer modeling of the disk.  In Section 5, we discuss the implications of the disk model and planet properties. In Section 6, we summarize our conclusions. 

\section{Observations and Data reduction}
\label{sec_obs}

\begin{table*}
\caption{Observational setup for PDS 70 imaging datasets.}
\label{tab_obs_pram00} 
\centering 
\small
\begin{tabular}{lclcrrrrccc} 
\hline\hline 
\multirow{2}{*}{Arm} & \multirow{2}{*}{UT Date} & \multirow{2}{*}{\makecell{Filter / \\ Prism}} & \multirow{2}{*}{DIT (s)} & \multirow{2}{*}{\makecell{Sci. \\  NDIT}} & \multirow{2}{*}{\makecell{Ref. \\  NDIT}} & \multirow{2}{*}{\makecell{Sci. \\  time(s)}} & \multirow{2}{*}{\makecell{Ref. \\ time(s)}} & \multirow{2}{*}{Seeing ('')} & \multirow{2}{*}{\makecell{Coherence \\ time (ms)}} & \multirow{2}{*}{\makecell{Wind \\ speed (m/s)}} \\ \\
\hline
IRDIS & July 15, 2021 & H-Pol & 16 & 80 & 32 & 1280 & 512 & 0.6 - 0.9 & 2.8 - 7.4 & 6 - 7 \\
IRDIS & Aug 21, 2021 & Ks & 16 & 136 & 80 & 2176 & 1280 & 0.6 - 1.5 & 4.6 - 15.2 & 0.2 - 4.4 \\
IFS & Aug 21, 2021 & YJH  & 32 & 68 & 40 & 2176 & 1280 & 0.6 - 1.5 & 4.6 - 15.2 & 0.2 - 4.4 \\
IRDIS & Aug 22, 2021 & Ks & 16 & 126 & 60 & 2016 & 960 & 0.6 - 0.8 & 4 - 8.9 & 3.3 - 4.8 \\
IFS & Aug 22, 2021 & YJH  & 32 & 68 & 30 & 2016 & 960 & 0.6 - 0.8 & 4 - 8.9 & 3.3 - 4.8 \\
IRDIS & Sep 02, 2021 & Ks & 16 & 26 & 16 & 416 & 256 & 0.6 - 1.0 & 4 - 7.6 & 5.6 - 6 \\
IRDIS & Sep 04, 2021 & Ks & 16 & 136 & 80 & 2176 & 1280 & 0.36 - 0.78 & 4.3 - 7.7 & 3.3 - 5 \\
IFS & Sep 04, 2021 & YJH  & 32 & 68 & 40 & 2176 & 1280 & 0.36 - 0.78 & 4.3 - 7.7 & 3.3 - 5 \\
IRDIS &	Feb 28, 2022 &	Ks	& 16	& 136	& 80	& 2176	& 1280	& 0.35 - 0.72 &	7.8 - 15.2 &	3.9 - 5.2 \\
IFS	& Feb 28, 2022	& YJH 	& 32	& 68	& 40	& 2176	& 1280	& 0.35 - 0.72 &	7.8 - 15.2	& 3.9 - 5.2 \\
\hline
\end{tabular}
\end{table*}

  \begin{figure*}
   \centering
   \includegraphics[width=\hsize]{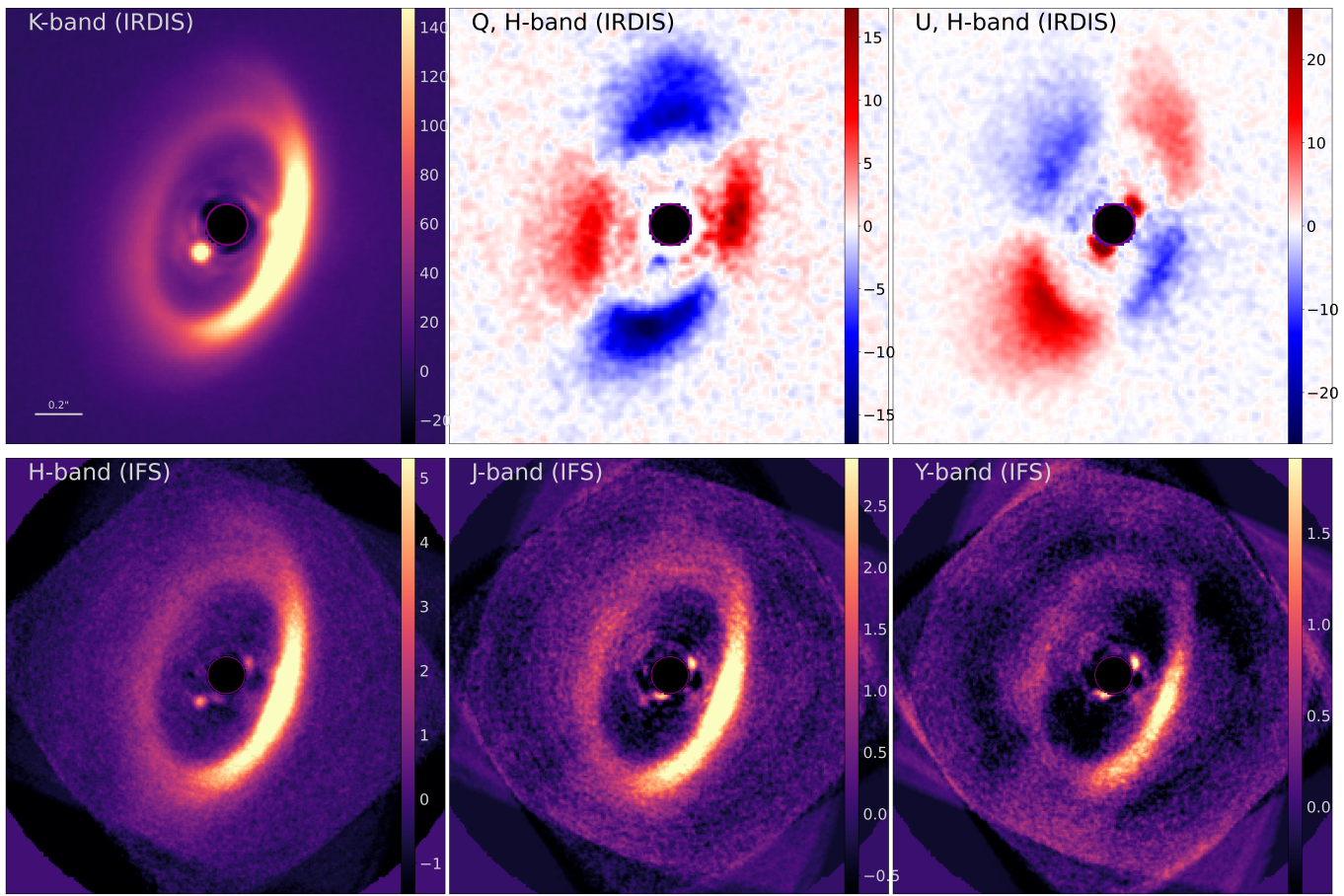}
      \caption{Top row, left to right: IRDIS $K_s$-band, Stokes Q/$H$-band and Stokes U/$H$-band images. Bottom row: $H$,$J$ and $Y$-band IFS observations, median combined over the typical band widths. North is up, east is left  and the black circle in the center is a mask over the coronagraphic region.}
         \label{fig_sphere_obs}
   \end{figure*}

The new observations presented in this paper were carried out using the SPHERE instrument \citep{2019A&A...631A.155B}, installed at the Nasmyth Focus of unit telescope 3 (UT3) at the VLT. SPHERE is a state-of-the-art high-contrast imager, polarimeter and spectrograph, designed to find and characterize exoplanets. It can deliver H-band Strehl ratios for bright stars (R$<$9) of up to 90\%, and continue to provide AO corrections for stars as faint as R$=$14. SPHERE also provides coronagraphs for starlight suppression, including apodized Lyot coronagraphs \citep{2005ApJ...618L.161S}. It is comprised of three subsystems: the infrared dual-band imager and spectrograph \citep[IRDIS;][]{2008SPIE.7014E..3LD}, an integral field spectrograph \citep[IFS;][]{2008SPIE.7014E..3EC} and the Zurich imaging polarimeter \citep[ZIMPOL;][]{2018A&A...619A...9S}.

We observed \prim\ in IRDIFS (IRDIS+IFS) \citep{2010MNRAS.407...71V,2014A&A...572A..85Z} and IRDIS-DPI modes \citep{deBoer:2020,vanHolstein:2020,vanHolstein:2017} on six nights between July 2021 and February 2022. Except for the July 2021 and the February 2022 observations, all were poor weather or partial data sets aborted because of worsening conditions. However, since we observed using star-hopping RDI \citep{Wahhaj:2021}, good reference point spread functions (PSF) could be subtracted to obtain useful data even for the aborted observations. This is because signal self-subtraction from insufficient sky rotation is not a problem, as it would be in Angular Differential Imaging (ADI). The observing parameters of the SPHERE observations are presented in Table~\ref{tab_obs_pram00}.  In summary, we observed \prim\ with IRDIS $H$-band obtaining Stokes Q and U, and total intensity images. We also obtained simultaneous IRDIS $K_s$-band imaging and IFS R=40 spectroscopy over the $YJH$-band region. We used the N\_ALC\_YJH\_S coronagraph, which has a mask radius of 93~mas and was designed for the $Y-K$ band range, with an overall transmission of 58\% \citep{2019A&A...631A.155B}.

\subsection{Total intensity}
 IRDIS  dual-band images in the $K_s$-band and IFS R$\sim$30-50 spectra (in the $Y-H$ range) are obtained simultaneously. The IRDIFS data were obtained in 1.5 h observing blocks (OBs), interleaving observations of PSF reference star with science observations. For the reference star, UCAC2 14412811 (R=11.7, H=8.9 vs R=11.7, H=8.8 for PDS 70), a 6 minute sequence was obtained after every 10 minutes of science, with only 1 minute taken to hop back and forth between the two targets. This star is separated 0.64$^o$ from \prim , and thus ideal as a PSF for reference differential imaging (RDI). This mode of RDI observations is called star-hopping \citep{Wahhaj:2021}.  

The star-hopping data are reduced according to the standard pipeline described in \citet{Wahhaj:2021}, with modifications to prioritize detecting extended emission over point sources. The details of the pipeline will be described in Swastik et al.\ 2024. In summary, the area used to match the science and reference PSF is chosen to be partly inside the coronagraph (separations 0.04$"$ to 0.08$"$), and  the region which includes the speckle ring right outside the AO control radius.
Since these are typically regions where circumstellar emission does not dominate, the disk emission is preserved beyond separations of 0.1$"$. For creating the best reference PSFs to subtract from each science image, we select the best 16 reference images based on the standard deviations of science--reference difference pairs. Then, the LOCI algorithm \citep{2007ApJ...660..770L} is used to create the best linear combination of reference images, minimizing the standard deviation in the difference images in the regions used for matching. The final reduced images are shown in Figure~\ref{fig_sphere_obs}.

\subsection{Polarimetry}
We used the IRDAP pipeline \citep{vanHolstein:2020} to reduce the $H$-band Dual-beam Polarimetric Imaging (DPI) mode data. IRDAP corrects for the instrumental polarization and polarization crosstalk, thereby attaining a polarimetric accuracy of $<$0.1\%. IRDAP subtracts the stellar polarization signal.
The reduced images are shown in Figure~\ref{fig_sphere_obs}. The $Q_\phi$ and $U_\phi$ images are shown in Appendix~\ref{app_qu_phi} for reference only, as we did not use them in our disk modeling efforts.
We remind the reader of a few key points on polarimetry and particular conventions used by the IRDAP pipeline, but for further details please review \citet{deBoer:2020}.  We measure polarization with respect to chosen axes: the North-South axis on sky denoted Q+, and the East-West axis denoted Q-. We chose axes (as opposed to directions) since the E-field  changes direction from positive to negative as a light wave propagates. The Stokes Q image is the difference in the time-averaged electric field intensity between the Q+ and Q- axes. The Stokes Q, U, and V images, which are practical to measure, represent intensity differences which together with the unpolarized light fraction constitute the total intensity of the observed light (see \citeauthor{deBoer:2020}~\citeyear{deBoer:2020}). The Radmc3D code, which we will use to perform radiative transfer modeling of the \prim\ disk, generates Stokes Q images with Q+ aligned along the horizontal axis, and North pointing right, East pointing up. Thus to match the IRDAP orientation, we rotate the Radmc3D images anti-clockwise by 90$^o$, so that Q+ is vertical, North is Up, and East is left.

\section{Planet Search Combining True-intensity Imaging and Polarimetry}
Angular Differential imaging (ADI) typically subtracts a significant fraction of circumstellar disk flux in a non-uniform fashion, as the science images themselves are used for PSF subtraction \citep{Marois:2006,Milli:2012}. This is facilitated by decoupling the pupil (thus PSF) and sky rotation, by controlling the telescope de-rotator. Thanks to star-hopping RDI, we have obtained a self-subtraction-free high-quality total intensity image of the disk in $H$-band. Such imaging obtained simultaneously with polarimetry is a first for PDS70. Other transition disks systems were recently observed in such a mode \citep{Ren:2023}. 

\begin{figure}
   \centering
   \includegraphics[width=\linewidth, trim=1.5cm 2cm 1cm 3cm, clip]{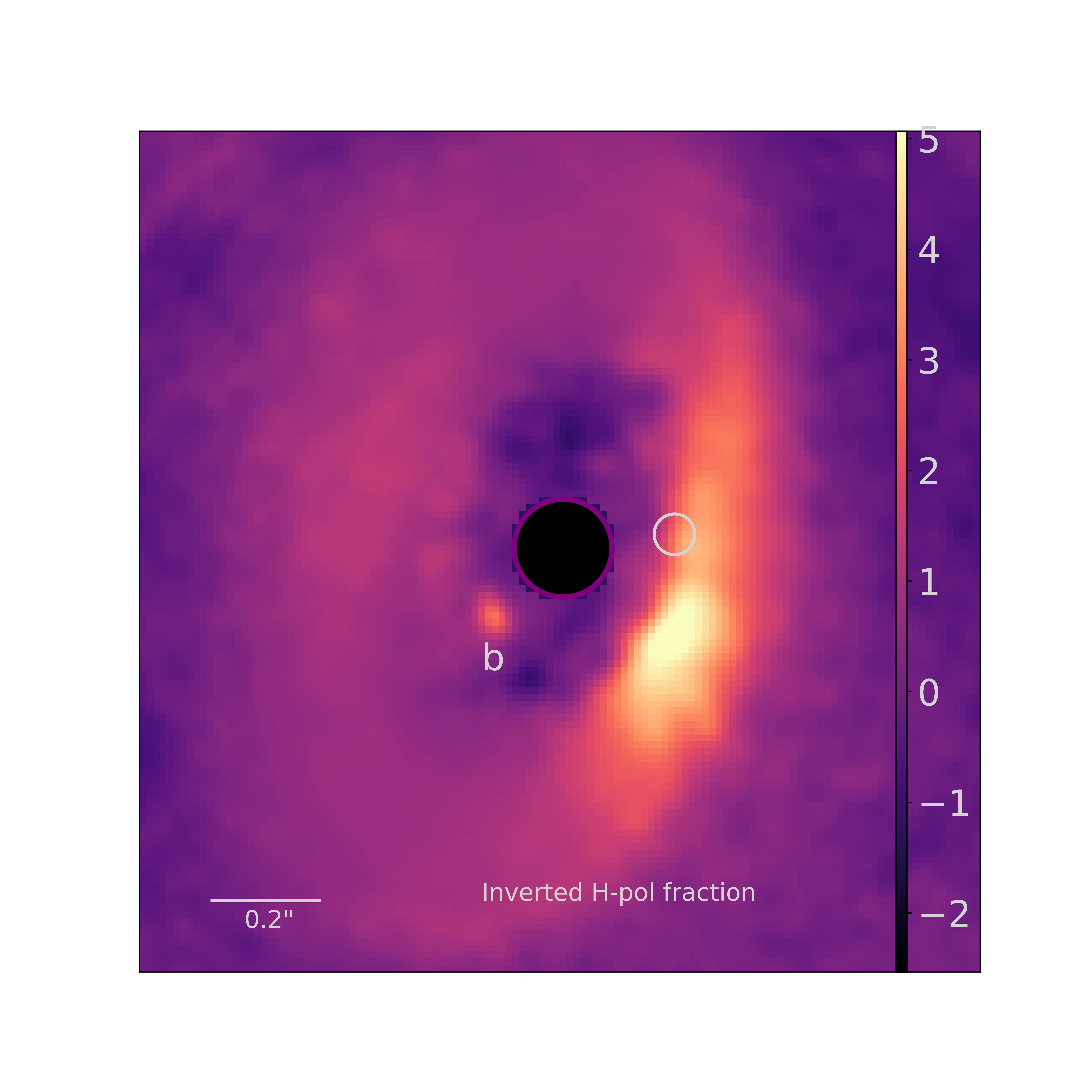}
      \caption{The inverted $H$-band polarized fraction map given by $I_{tot}/\sqrt{Q^2+U^2}$, which could reveal low-polarization planets superposed on a highly polarized, high SNR disk. No new planets are detected. Planet $b$ seems well-recovered, but planet $c$ is not detected (empty circle). The reason is explained in the text.}
         \label{fig_pfac_inv_img_hband}
   \end{figure}

Our novel observational method allows us to search for planets in a new way.
Planet atmospheres should scatter light with low polarisation fractions, while the disks are expected to scatter with much higher polarization \citep{Stolker:2017}. Thus, the ratio of the linearly polarized intensity $\sqrt{Q^2+U^2}$ and the total intensity, could reveal planets as dips or negative Gaussians superposed on a bright polarization fraction map. We will use the inverse map, so that planets show up as bright spots. Such a map should only be trusted in the regions where the denominator has high signal-to-noise ratio (SNR), i.e., not close to zero.  Thus, when the SNR falls below 5, the polarized intensities are set to the constant value corresponding to an SNR of 5. The inverse map, $I_{tot}/\sqrt{Q^2+U^2}$, which should show planets as bright spots, is shown in Figure~\ref{fig_pfac_inv_img_hband}. Planet $b$ is clearly detected, but not planet $c$. The method fails for planet $c$ as $\sqrt{Q^2+U^2}$ has very low SNR at that location and thus provides a flat divisor, so that there is no signal boost for planet $c$. Another problem is that planet $b$ and the front part of the disk appear comparably bright in the map, indicating both have low-polarized fractions. Thus, we conclude that this method would only be useful in the case of a low-polarization planet superposed on a high-polarization disk detected at high SNR.

\subsection{Improved planet extraction in disk systems}

\begin{figure}
   \centering
   \includegraphics[width=\linewidth, trim=1.5cm 2cm 1cm 3cm, clip]{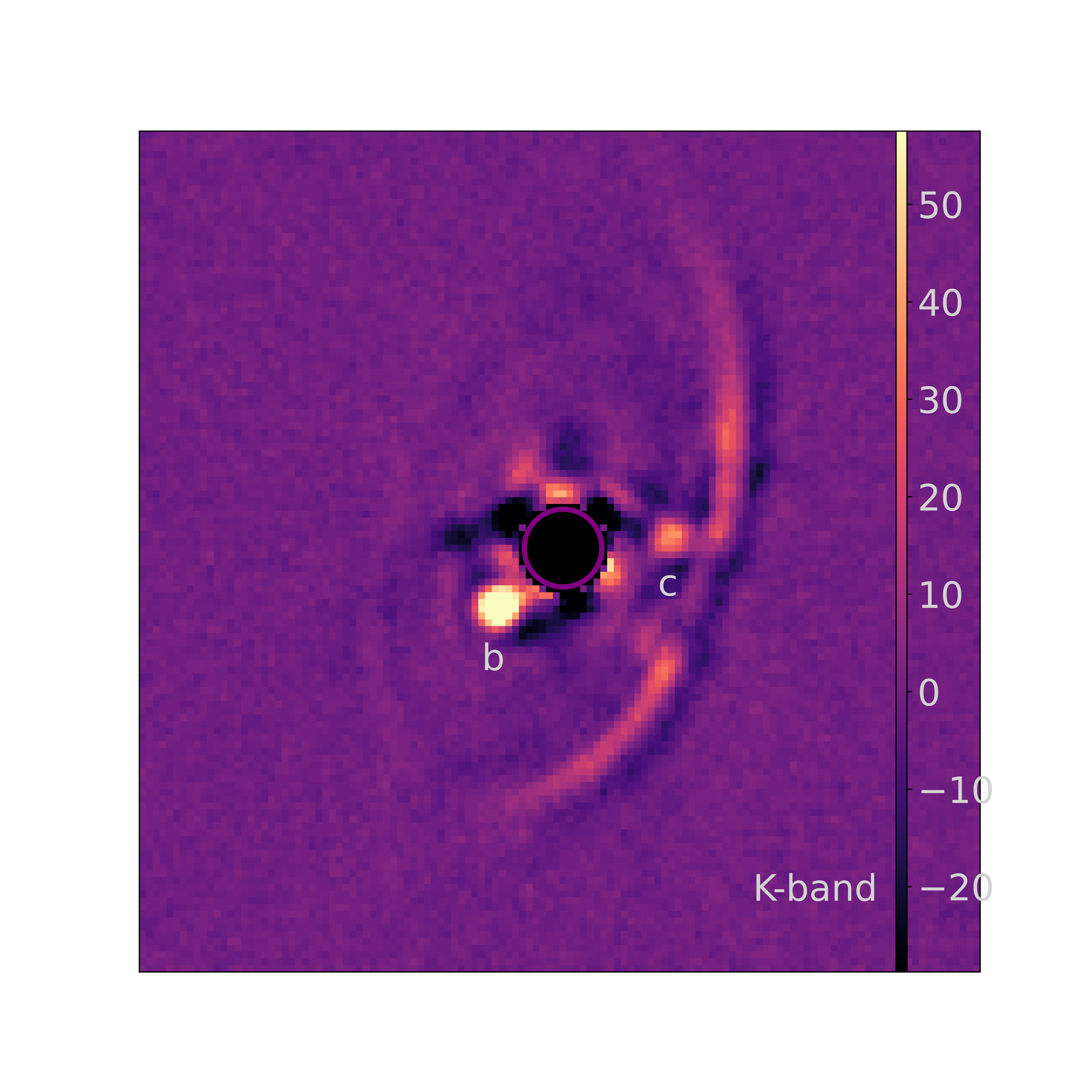}
      \caption{IRDIS $K_s$-band reduction to bring out the planets, by trying to remove the disk. A smoothed version of the disk reduction (median smoothed over 12 pixel box) along with the PSFs of the reference star was supplied to the LOCI algorithm to subtract both disk and star.}
         \label{fig_planets_rm_disk}
   \end{figure}

Detecting planets is more difficult when imaging bright disk systems. The disk emission is so strong in the PDS 70 NIR images that it interferes with the PSF subtraction. Thus, we attempt to subtract a smoothed version of the disk, along with the subtraction of the reference star's PSFs. This time LOCI is used to match the PSFs over the  0.1$''$--1.5$''$ annulus centered on the star, instead of trying to avoid the disk regions as in the first-round reduction (see section~\ref{sec_obs}). The smoothed disk image is made by median smoothing the disk image we obtained in the first-round reduction, over a box size of 12 pixels. The median smoothing removes most of the point sources and sharp features in the disk. Thus removing the smoothed disk from the data should not remove candidate companions. This works better than unsharp-masking the images prior to reduction, because unsharp-masking leaves strong artifacts from the disk in the images. In Figure~\ref{fig_planets_rm_disk}, we see the first unambiguous detection of planet $c$ in the NIR. We measure of the location of the planet using the IDL $bscentrd$ routine, a very robust routine centroiding routine we tested extensively in \citet{Wahhaj:2013}. We estimate the SNR of planets $b$ and $c$ to be 11.5 and 5.5, respectively. It is also clear that there are no other planets as bright as planet $c$ detected in the image, beyond a separation of 0.1$"$ from the star (beyond the coronagraph). It is possible that the disk is hiding a brighter planet or CPD than planet $c$. Planet $c$ clearly lies in the gap, just inside the inner edge of the disk. This reduction demonstrates the effectiveness of subtracting a smoothed disk along with the reference star PSFs, in isolating and detecting point sources. Planet $c$ was hardly noticeable in the reduced image of the disk presented earlier (see Figure~\ref{fig_comp_mod1}). The FWHM of both planet detections are 59~mas ($\sim$7\,au) very close to the $K_s$-band diffraction limit, indicating that any CPDs have compact emissions originating from close to the planets.

\begin{figure}
   \centering
   \includegraphics[width=\hsize]{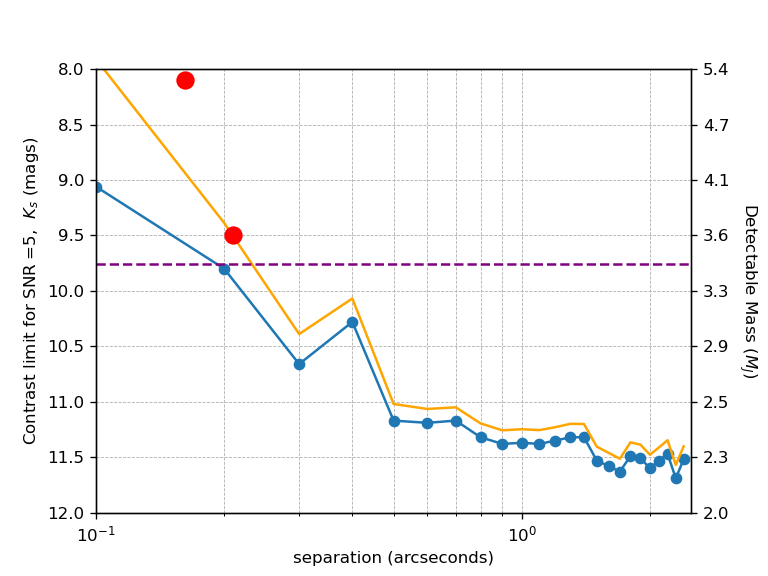}
      \caption{IRDIS $K_s$-band detection contrast-limit for SNR $>$ 5 shown as the blue curve. It was made from simulated planet recovery as explained in the main text. The orange line shows the limits corrected for small sample statistics according to \citep{mawet2014}.The mass detection limits are estimated from the DUSTY models, assuming an age of 5~Myrs for PDS~70. The dashed line represents the detection limit at the brightest part of the disk.}
         \label{fig_kband-contrast-curve}
   \end{figure}

The $K_s$-band image reveals the PDS~70 circumstellar environment beyond 0.1$"$ (the coronagraph radius). To find the detection limits as a function of separation, we calculate a simple 5$\sigma$ contrast curve as $5\sigma(r)/(estimated\ stellar\ peak)$, where $\sigma(r)$ is the robust standard deviation as a function of separation. The stellar peak is estimated from the background object to the North \citep{Riaud:2006}, Hashimoto et al.\ 2012), which had $PA$ = 14.1$^o$, separation = 2.62$''$, $\Delta$K = 4.58 mag in our non-coronagraphic short-exposure images, which detected both stars without saturation. The background object is also well-detected yet unsaturated in all our science images. We take the robust $\sigma(r)$ to reduce the disk's effect, which is already attenuated by removing the smoothed disk, as mentioned above.  However, we indicate in Figure~\ref{fig_kband-contrast-curve} the detection-limit on the brightest part of the disk, 3.4\mjup. Thus planets of that mass could be obscured where disk detection is strong.

\begin{figure}
   \centering
   \includegraphics[width=\hsize, trim=2cm 2cm 2cm 2cm, clip]{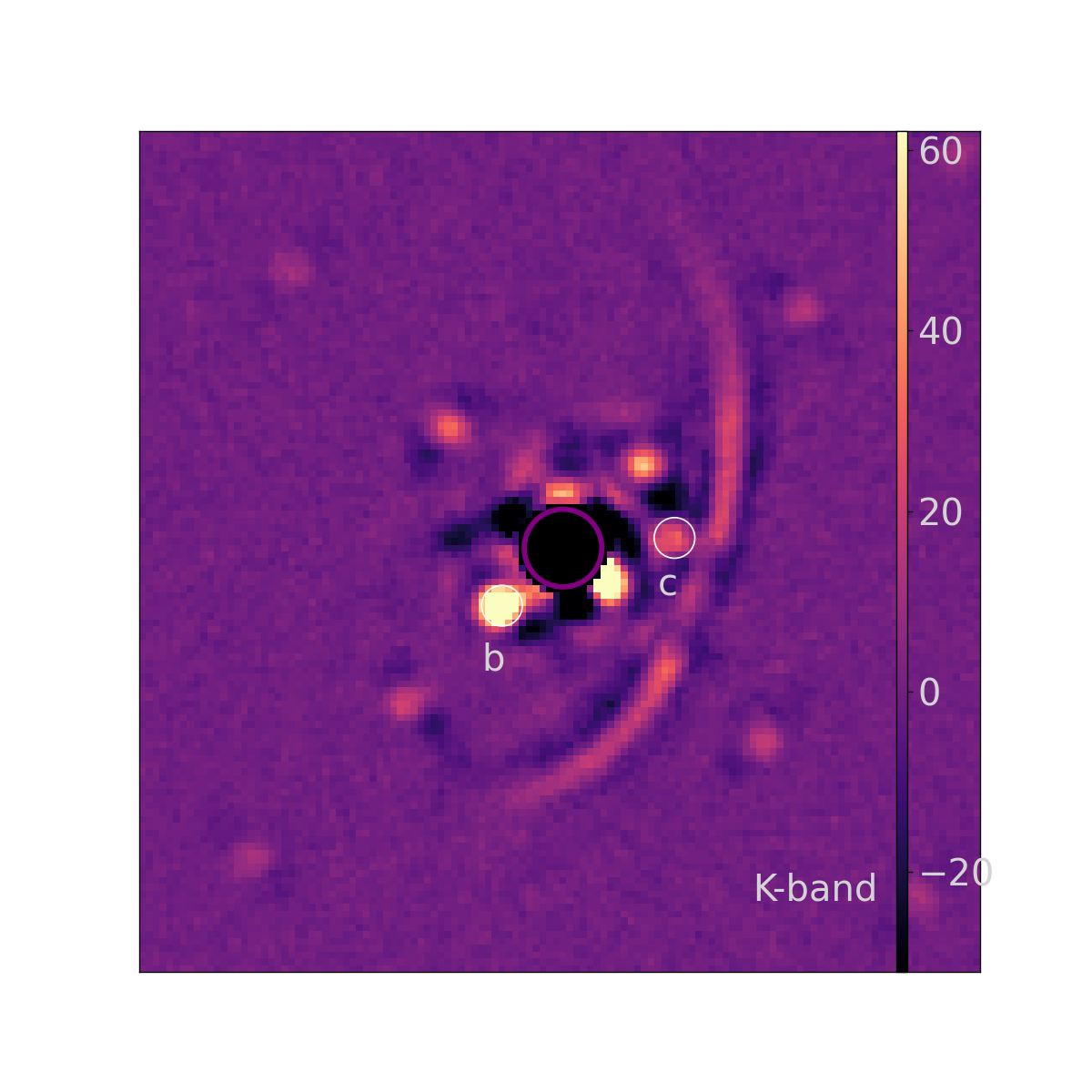}
      \caption{IRDIS K$_s$ reduction to recover simulated planets, and to detection limits as shown in Figure~\ref{fig_kband-contrast-curve}. The disk has been subtracted as described in the text. The real planets $b$ and $c$ are labeled.}
         \label{fig_kband-sim-planet-recov}
   \end{figure}

Given the complexity introduced by the disk residuals and the two planets, in the smooth-disk subtracted reduction, improved contrast estimation by recovery of simulated planets is difficult. Nevertheless, we injected simulated planets in the more cleared regions of the system, into the basic reduced images, at twice the fluxes of the detection limits. We started the injections started at a PA of 225$^o$ and 0.1$''$ separation, incrementing the polar coordinates by 0.1$''$ and 90$^o$ for each subsequent planet. A first round reduction of this data set is achieved exactly as before, trying to preserve the disk. Again, a smoothed version of the reduced image, this time with the simulated planets and the disk, is supplied as a reference image for a second rounds of LOCI reduction. This is basically a repeat of the smooth disk-subtracted reduction. The resulting image is shown in Figure~\ref{fig_kband-sim-planet-recov}. Since there are few clear regions in the disk, the flux uncertainty as a function of separation is estimated from this single reduction only. The detections of the simulated planets passed criteria based on SNR and shape discussed in \citet{Wahhaj:2013}. We take twice the difference between the actual flux  of a simulated planet (in a 2-pixel radius aperture) and the recovered flux as the uncertainty, although this would normally be considered the systematic error and not the random error. Thus, we estimate the uncertainties as 0.26~mag at 0.1$''$ and 0.12~mag at larger separations. The mass estimates resulting from this photometry, assuming an age of 5$\pm$1 Myrs \citep{Muller:2018} and hot start models \citep{2003A&A...402..701B} are presented in Table~\ref{tab_planet_props}.

We reach an unprecedented detection limit of 5\mjup~at 0.1$"$ separation. Compared to past observations, in our novel detection space of $\sim$11 to 22~au, we find no new planets. This improved sensitivity to planets is due to better stellar PSF subtraction, thanks to star-hopping RDI and the smoothed disk subtraction, as opposed to what is achievable with just ADI, especially in the presence of disk signals. The deepest contrast is reached at $\sim$90~au separation, where the detection limit is 2.3\mjup .

\begin{table}[h]
\centering
\begin{tabular}{c c c c}
\hline
\hline
\multicolumn{1}{c}{Planet} & \multicolumn{1}{c}{$dK_{\text{mag}}$} & \multicolumn{1}{c}{Mass} & \multicolumn{1}{c}{Epoch} \\
\hline
$b$ & 8.1$\pm$0.3 & 5.4$\pm$0.5\mjup & 28/2/2022 \\
$c$ & 9.5$\pm$0.1 & 3.8$\pm$0.4\mjup & 28/2/2022 \\
\hline
\end{tabular}
\caption{Planet Properties}
\label{tab_planet_props}
\end{table}

\subsection{Spectra extraction}

\begin{figure*}
   \centering
   \begin{tabular}{c}
   \includegraphics[width=0.8\textwidth]{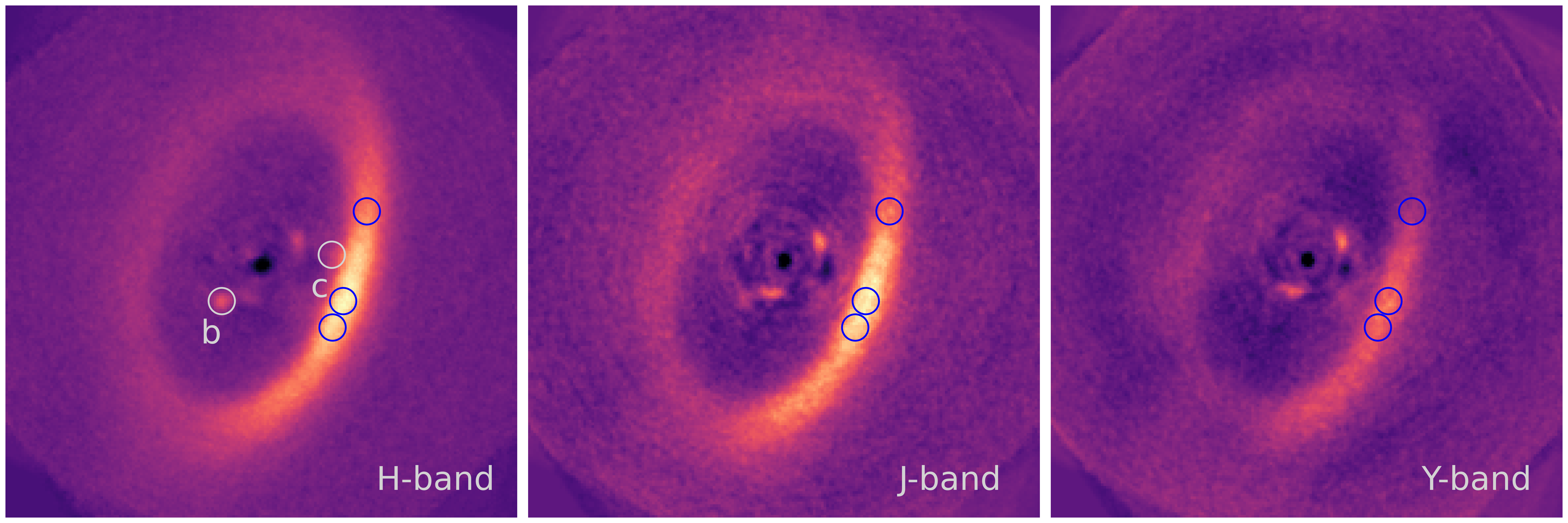}\\
   \includegraphics[width=0.6\textwidth]{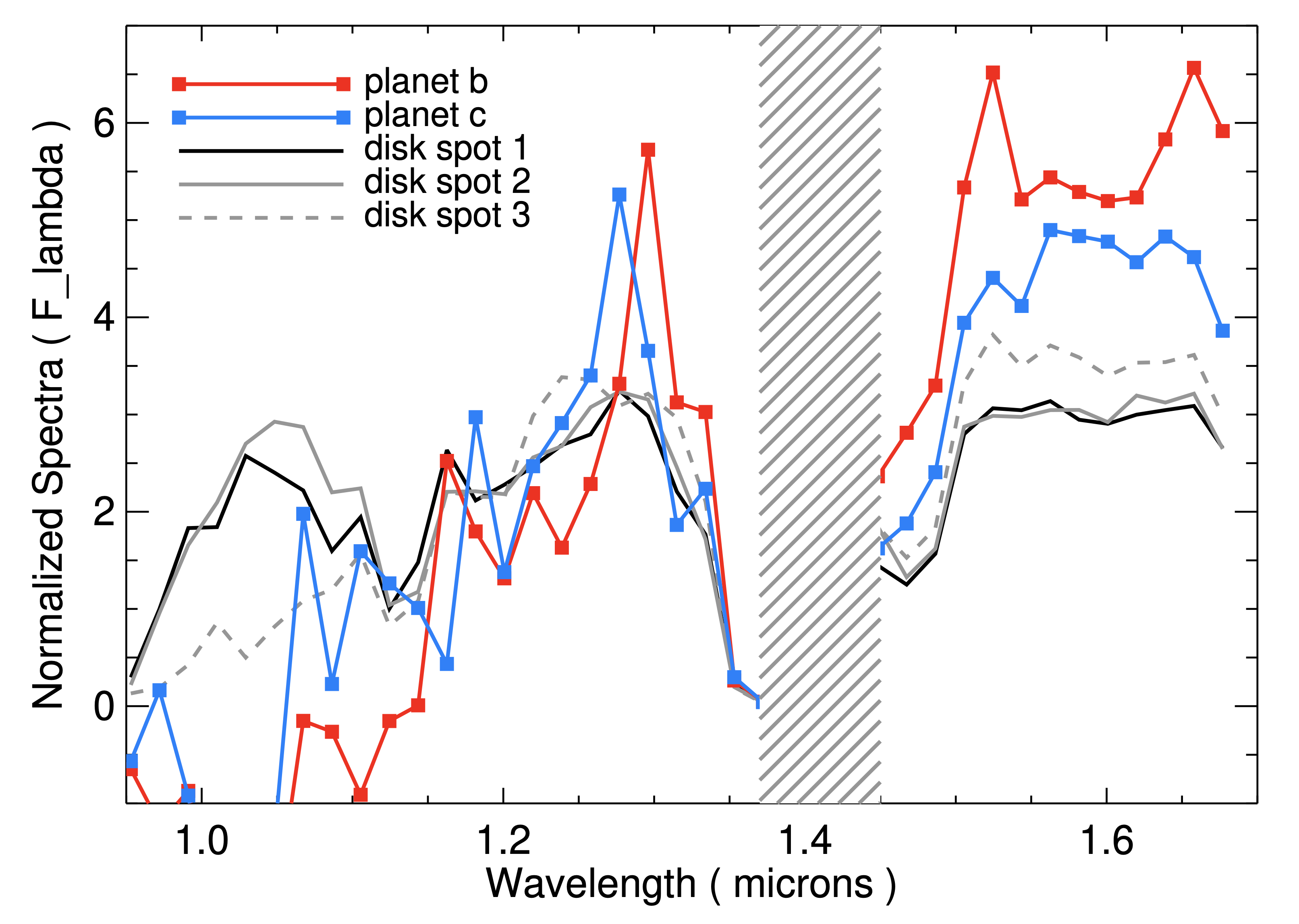}
   \end{tabular}
      \caption{Top: $YJH$-band images made from collapsing the IFS spectra, and the locations from which the planet and disk spectra were extracted. Bottom: IFS Spectra of the $b$ and $c$ planets, along with spectra of this dust ring at three locations. The planet spectra are similar to each other, and clearly different from the disk spectra. The region contaminated by telluric absorptions around 1.4\mic\ is grayed out.}
         \label{yjh-disk-planet-spots-ifs}
   \end{figure*}

We can extract the planet spectra from the 39 $Y$--$H$-band IFS channels, at the locations of the planet detections in the $K_s$-band image. We use an aperture of radius 3 spaxels (7.46$\times$7.46~mas) to extract flux per IFS channel at five spots: two on the planets, and three on selected bright parts of the disk (see Figure~\ref{yjh-disk-planet-spots-ifs}). These IFS channel fluxes are divided by the stars' channel fluxes, obtained from the non-coronagraphic $FLUX$ observations. To obtain the final spectra in Figure~\ref{yjh-disk-planet-spots-ifs}, we multiply these channel flux ratios by the stars' actual spectra (obtained with the XSHOOTER instrument by \cite{Campbell-White:2023}).

As in the IRDIS reduction, we estimate the flux uncertainty in each spectral bin to be 0.25~mag, which indicates that the uncertainly in the first order slope of the spectra is $\sim$4\% (0.25~mag /$\sqrt{\text{number of spectral channels}}$.) Thus, it seems that the planet $b$ and $c$ spectra are quite similar to each other, and are clearly redder than the disk spectra over the $YJH$-bands. Particularly, the planet spectra peak around 1.27 \mic\ and between 1.5 and 1.67 \mic , while they are clearly suppressed between 0.9 and 1.15 \mic , relative to the disk. The $J$ and $H$-band features are likely due to methane absorption (see Pluto's spectra in Grundy et al.\ 2013). The disk spectra along different lines of sight are also very similar. One may argue that the aperture flux obtained at planet $c$ location has disk contributions to it, and also that our planet $b$ spectrum does not have proper background subtraction. But the fact that we can differentiate between planet and disk spectra despite these potential systematic biases indicates that the differences in the spectra are likely real. 

The main reason for the difference between disk and planet spectra is that the planet atmospheres are around 1100~K \citep{Wang:2021}, whereas the disk is much colder . The evidence for the CPD comes from the H$_\alpha$ accretion signature  \citep{Haffert:2019} and the ALMA 855\mic\ detection \citep{Isella:2019}. In contrast, the $JHK$-band spectra seem to originate from the planetary atmospheres and not from the CPDs. The spectra also suggest a good way to distinguish between planets and disk material.

\subsection{Orbit fitting}
\label{sec_orbit_fitting}
We use the astrometry collated by \citet{Benisty:2021}, along with those from our epochs (see Table~3) to find constraints on the orbits of planets $b$ and $c$. Planet $c$'s orbit is less well known than planet $b$'s as it has been detected at fewer epochs, and with less SNR. We use the $orbitize!$ Python package written by \citet{Blunt:2017} to find the most likely orbital solutions. 

\begin{figure*}[ht!]
   \centering
   \begin{tabular}{cc}
   \includegraphics[width=0.45\textwidth]{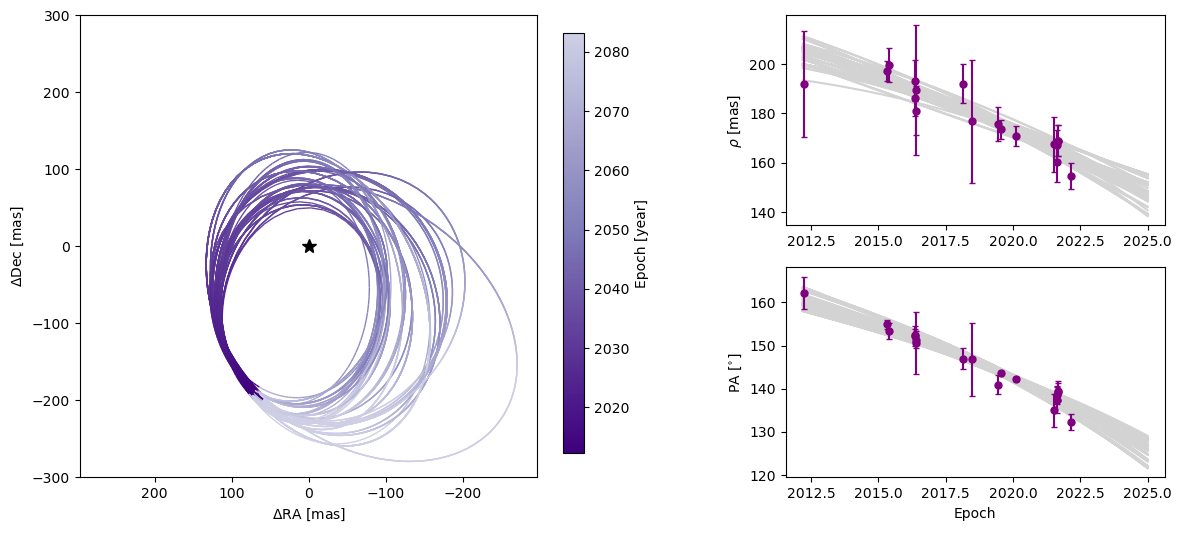}&
    \includegraphics[width=0.45\textwidth]{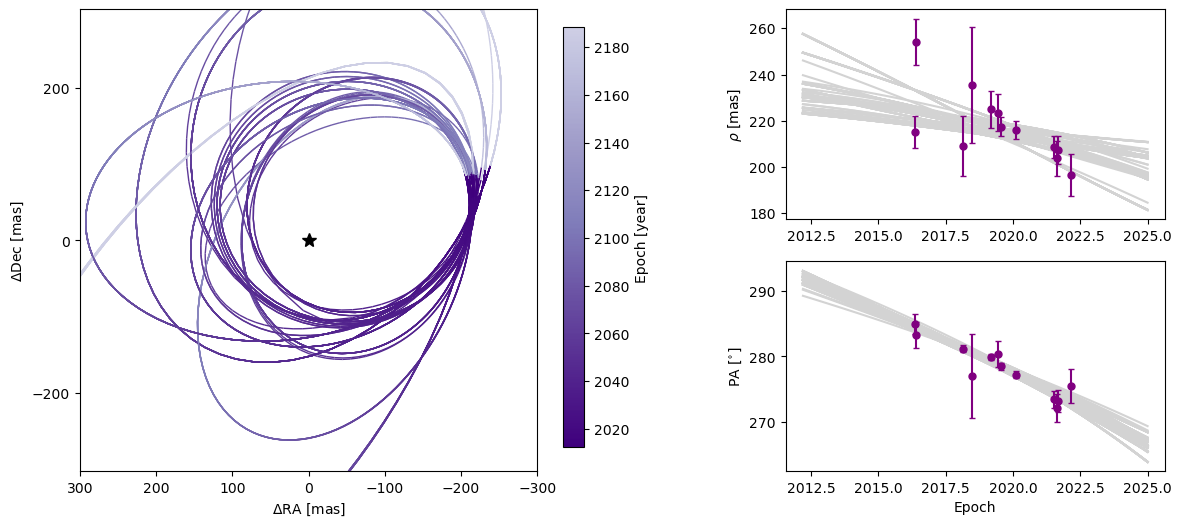}
    \end{tabular}
    \caption{Orbit fitting using astrometry from this work and earlier studies for planet $b$ (left) and $c$ (right).  The best 500 orbits from 1,000,000 orbits fitted by the $orbitize!$ package \citep{Blunt:2017}, shown along with predictions of separation and PA for the near future.
      }
         \label{fig_orbit_planet_b_and_c}
   \end{figure*}

We use the package's MCMC method, sampling 1,000,000 orbits. The results are shown in Figures~\ref{fig_orbit_planet_b_and_c}, ~\ref{fig_orbit_planet_b_corner} \& ~\ref{fig_orbits_c_corner_plot}. 
For both planets, the eccentricities are likely $>$0.2 and $<$0.7, favoring significantly eccentric orbits.
Planet $b$'s orbital constraints are semi-major-axis (SMA) $= 15.7^{+2.3}_{-1.4}$~au, eccentricity $=$ 0.46$^{+0.08}_{-0.1}$ and inclination $=$ 49.3$^{o+23^o}_{-14^o}$ (subtracting from $180^o$), consistent with that of the disk.
Planet $c$'s orbital constraints are semi-major-axis (SMA) $= 17.6^{+2.9}_{-1.5}$~au, eccentricity $=$ 0.47$^{+0.09}_{-0.12}$ and inclination $=$ 58.5$^{o+24^o}_{-20^o}$ (subtracting from $180^o$), again consistent with that of the disk.

We note that 2:1 mean-motion resonance (MMR) orbits are allowed by our constraints, for example, if we assume SMA of 14 and 22.2~au for planets $b$ and $c$ respectively. These SMA lie within a 75\% confidence interval in our constraints. Thus, reasonably stable orbits should allowed by our constraints. An SMA of 22.2~au corresponds to a maximum separation of 32.6~au (=SMA[1+e]) from the star, which would mean that planet $c$ orbits far away from the gap edge at 49.7~au. Since we can now easily extract planet $c$'s astrometry using our new RDI with disk subtraction technique (see Section~\ref{sec_obs}), better orbital fits should be forthcoming in the near future.

Our constraints are somewhat more relaxed than those found by \citet{Wang:2021} as we do not impose any additional constraints on the orbits. Quite justifiably, \citet{Wang:2021} imposed requirements that the orbits should be co-planar, stable for 8~Myrs, and not have peri-astrons that exchange order over their lifetimes. We also found the orbital constraints using only the astrometry from \citet{Wang:2021}, and these were quite similar to the ones presented here.  Thus, we note that we find higher eccentricities for both orbits, not because of our new astrometric points, but because we do not impose the stability conditions of \citet{Wang:2021}. Given the significant eccentricities, the planetary orbits may become unstable given another 5~Myr.  

\begin{table}[h]
\label{tab_new_astrom01}
\centering
\caption{New Astrometry for planets $b$ and $c$. Epochs are in Modified Julien Dates (MJD).} 
\begin{tabular}{cccccc}
\hline
\hline
MJD & Planet & Sep. (mas) & $\pm$ & PA ($^o$) &$\pm$ \\
\hline
59410 & b & 167.5 & 11 & 135.0 & 4 \\
59447 & b & 160.3 & 8 & 137.4 & 3 \\
59448 & b & 167.2 & 6 & 138.5 & 2 \\
59459 & b & 169.0 & 6 & 139.5 & 2 \\
59461 & b & 168.9 & 6 & 139.3 & 2 \\
59638 & b & 154.6 & 5 & 132.3 & 2 \\
\hline
59410 & c & 208.6 & 5 & 273.4 & 1 \\
59448 & c & 203.7 & 8 & 272.1 & 2 \\
59459 & c & 207.3 & 6 & 273.2 & 2 \\
59638 & c & 196.5 & 9 & 275.5 & 3 \\
\hline
\end{tabular}
\end{table}

\section{Radiative Transfer Disk Modeling}

In this section, we model the three main datasets: total intensity K$-$band imaging, polarimetric H$-$band imaging, both of which are from the SPHERE instrument, and 870$\mu$m imaging from ALMA \citep{Benisty:2021}. We produce models of a flared protoplanetary disk of  gas and dust, using the radiative transfer code Radmc3D (Dullemond et al.\ 2006). Given that each model takes about one minute to produce, and given the high information content of our datasets, we do not attempt to converge on a true best fit model. Rather, we attempt to find a model that is broadly consistent with the data given our computational constraints. We also study the constraints on our parameters by exploring the \chisq\ space around the best-fit model.  Thus we find a range of models that are broadly consistent with the data, but do not claim that these are formal constraints. 

\subsection{Search for a convergent model}

Given that we had to obtain a good model in a reasonable amount of time, the choice of the metric to minimize was important. Ordinarily, we would choose the total $\chi^2$ metric, adding the $\chi^2$ for the three images and three flux measurements. But using the total $\chi^2$ metric would force search algorithms to prioritize minimizing the $\chi^2$ for the highest resolution image, since it contributes the most to the total $\chi^2$. If the search has not yet converged on the actual minimum, the current best model may poorly fit the smaller data components, such as the band photometry and low-resolution images. Instead, we use the average reduced $\chi^2$ of all the images and photometry, so that they have equal weight in the combined metric. For more details, see Appendix~\ref{app_goodfit}. We warn that our model searches are only locally convergent, but may be missing the global minimum. However, since we repeated the searches many times, exploring reasonable guesses at improved initial parameter values, better solutions would be very difficult without considerably more computing power.

\subsection{Model parameters}
\label{sec_mod_prams}
\begin{figure}
   \centering
   \includegraphics[width=\hsize]{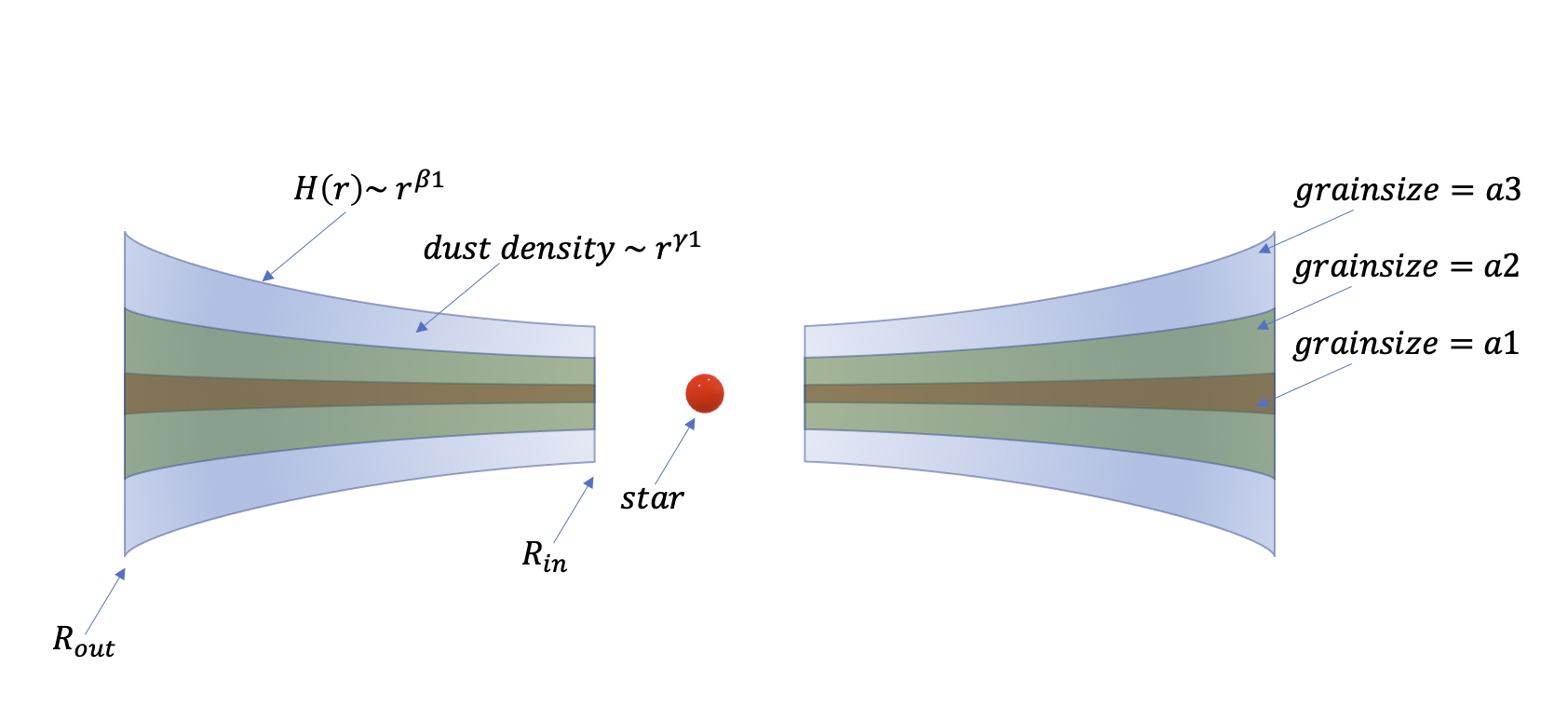}
      \caption{A schematic of the model dust disk as described in section~\ref{sec_mod_prams}.}
         \label{fig_disk_schematic}
   \end{figure}

Our first order model is chosen to be a flared disk of gas and dust extending from $\sim$0.1 au to $~$100au, with a gap inside this disk (see schematic in Figure~\ref{fig_disk_schematic}). This model is inspired by the findings of previous studies on PDS~70 \citep{keppler:2018, Portilla-Revelo:2022, Benisty:2021}, and our own $K_s$-band image. On a cursory examination of our data, we note that the ALMA image shows a flat disk, while the SPHERE NIR images show a flared-disk. This indicates a vertical settling of large grains (mm-sized). Thus, we will use multiple populations of grains which are allowed to spatially segregate. Each grain population is said to constitute a disk zone. Initially, we attempt to fit the data with a 2-zone model and later complexify it with a third zone.
In the NIR images, we also notice that the disk is much brighter in the southwest (SW) direction, which suggests significant forward scattering. This portends the existence of micron-sized grains. However, since the northeast (NE) side is also well-detected, there must be sufficient back-scattering grains, especially in the polarimetric images. According to our modeling experience, this indicates the presence of sub-micron grains.

Our radmc3D models are made at low-resolution for computational speed as we describe in appendix B. However, once we have converged on a best-fit model, we re-run the models at higher resolution to check if there are vast changes in fit quality. Since this is not the case, we consider the resolution of our models to be adequate. We note here that the model images are convolved with Gaussians of width equivalent to the PSFs of the observations.

Each disk zone is described by 1) a grain size, $a$, 2) an inner radius \rin, 3) an outer radius \rout, 4) a scale height \hrr $= H_0 (r/R_{in} )^{\beta}$, where $r$ is the stellocentric distance 5) a dust density power-law $r^{-\gamma}$, 6) a disk mass \mdisk, 7) a position angle $PA$ and 8) an inclination angle $i$ wrt. to the plane of the sky. Most of these parameters are common to all disk zones. However, some zone parameters are allowed to be independent, specifically grain sizes, zone disk masses, scale height and \plsig. We note here that $R_{in}$ and $R_{out}$ are the same for all the zones. Also, between the star and $R_{in}$ there is a gap which has only a fraction (e.g. 1\%) of the exterior dust density. We call this fraction {\it gap clearing}.  The dust grains were modeled as amorphous olivine with 50\% Mg and 50\% Fe, or pyroxene grains with 70\% Mg and 30\% Fe, both of which are used often in the literature \citep{1994A&A...292..641J,1995A&A...300..503D}. We used the Powell algorithm \citep{1964Powell} to improve the fit, after several iterations of parameter estimation by eye, which proved to be much more efficient. The Powell algorithm is known to be robust and efficient for minimization over many variables, when the gradient of the optimization metric is not known a priori. To calculate \chisq , we follow the method of \citet{2005ApJ...618..385W}, where the \chisq\ are normalized according to the number of significant resolution elements they represent (see details in Appendix ~\ref{app_goodfit}).  

\subsection{Best 2-zone model}
\label{sec_mod_fit_res}
\begin{figure}
   \centering
   \includegraphics[width=\hsize]{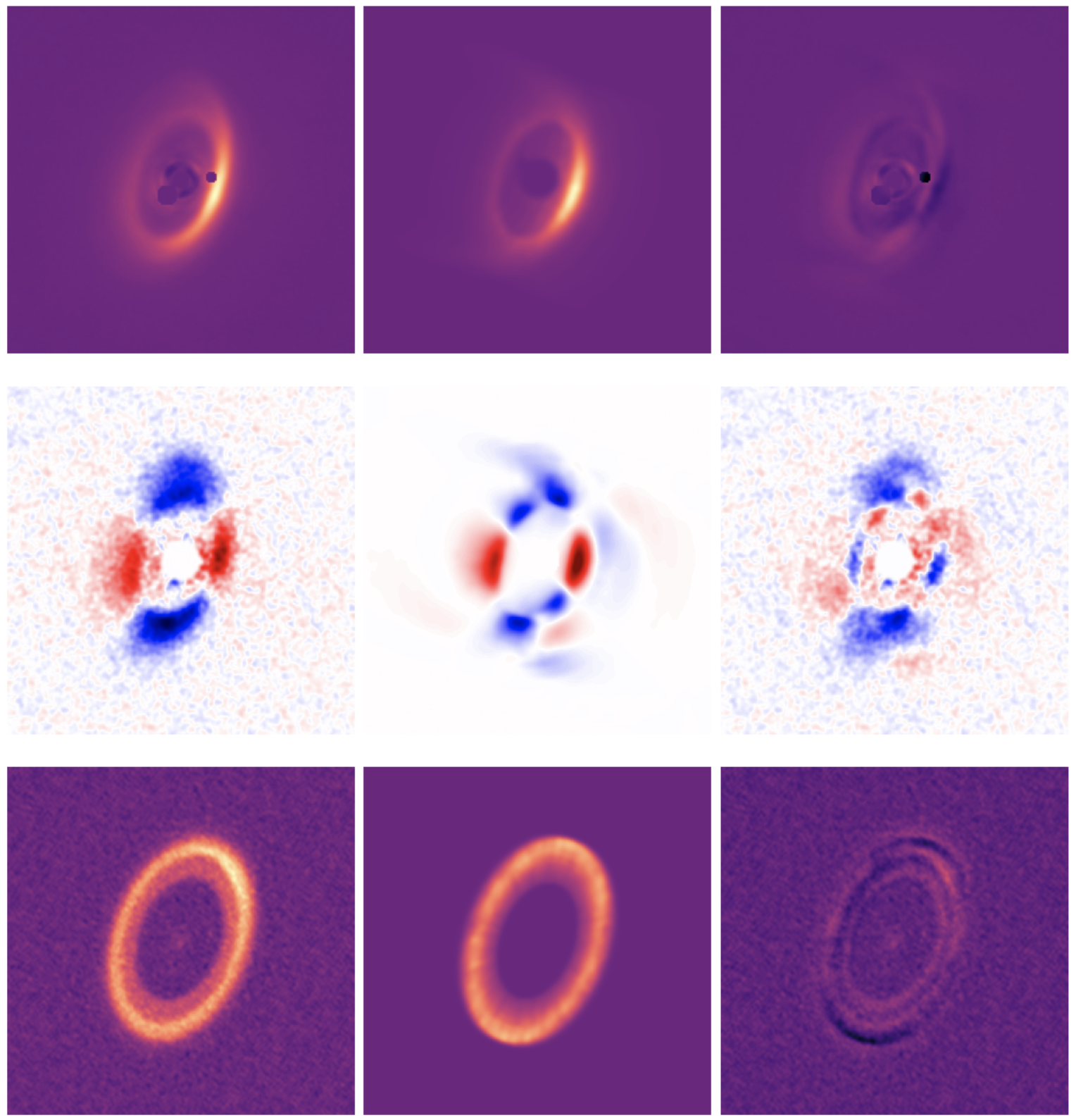}
      \caption{Comparison of Radmc3D preliminary 2-zone model images with data, showing poor fits to the polarimetry data. Left, middle and right columns are data, model, and residual respectively. The top, middle and bottom rows are $K_s$-band total intensity image, $H$-band Stokes Q image, and ALMA 870\mic image respectively. The data and model images are of the same resolution.}
         \label{fig_comp_mod1}
   \end{figure}

The best-fit 2-zone model, that has 14 free parameters, is compared to the observed images and residuals after differencing in Figure~\ref{fig_comp_mod1}. The model parameters are shown in Table~\ref{tab_mod1_prams}. Qualitatively, all the salient features in the data seem to be present in the 2-zone model: a deep gap~50~au in the inner disk, the polarimetric model image shows an optically-thick flared disk, the \submm\ model shows a flat disk, the NIR total intensity model shows an outer arc NW of the disk, and strong forward-scattering, while the polarimetric model images also show some back-scattering. The flux comparisons yield reduced \chisq s close to one, indicating good matches. However, several short-comings of the model are already evident to the eye, as we discuss below. Moreover, the image model comparisons yield reduced \chisq values that indicate poor fits (see  Table~\ref{tab_gof_stats01}), despite the fact that most of the disk flux has been subtracted. This is not surprising since the observations are of such high signal-to-noise and resolution that they warrant highly complex models with $\sim$ 100 parameters, given that our observations represent $>$ 2000 data points ( $>$ 1000 resolution elements with high SNR; see Table~\ref{tab_gof_stats01} and Appendix~\ref{app_goodfit}). 

A clear deficiency of the 2-zone polarimetric model is insufficient disk emission from the Northern and Southern parts of the disk (which indicates deficit of light with E-field oscillating East-West). From our modeling experience, this deficit seems unavoidable for grain sizes $\sim$ 0.5--1\mic, where the grains have a non-trivial polarimetric phase function. Grains above this size are too forward-scattering to satisfy the polarimetric image, while grains below this size do not forward-scatter enough to satisfy the total intensity NIR image. The disk flaring required for the micron-sized zone also seems unrealistically high (13\% at \rin\ and 150\% at \rout ). The 870\mic\ model flux is double the data flux, suggesting that we need grains with higher albedo to scatter more light in the NIR and produce less thermal emission in the \submm . Pyroxene grains may be used to compensate for this limitation. Thus, we experiment with adding a third disk zone with sub-micron grains, and allowing pyroxene grains where helpful next.

\subsection{Best 3-zone model}

\begin{figure*}
\centering
\includegraphics[width=0.9\textwidth]{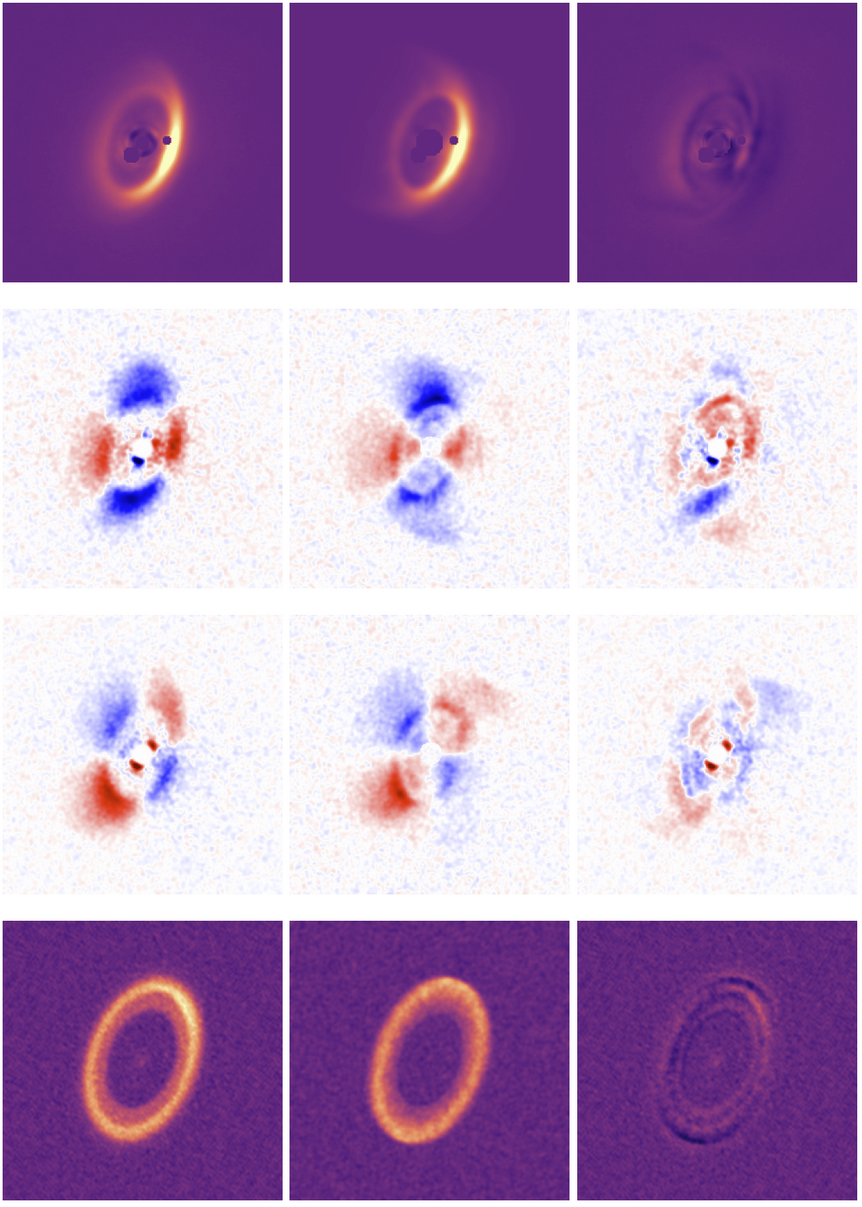}
\caption{Comparison of Radmc3D 3-zone model images with data. Left, middle and right columns are data, model, and residual respectively. The rows from top to bottom are: $K_s$-band total intensity image, $H$-band Stokes Q image, Stokes U image, and ALMA 870\mic image respectively.}
\label{fig_comp_mod2}
\end{figure*}

For the 3-zone model, we allowed the search algorithm to pick between olivine or pyroxene grains for each zone. Also, Zone1 and Zone2 are forced to have the same properties except for dust mass and grain size. Zone3 is allowed to have a different dust mass, grain size, $\beta$, $H$ and $\gamma$. In summary, we now have 17 free parameters compared to 14 for the 2-zone model. The model parameters are shown in Table~\ref{tab_mod2_prams}. The best-fit 3-zone model is compared to the observed images in Figure~\ref{fig_comp_mod2}. The flux in the \submm\ is now very well-matched by the model. The disk flaring for the micron and submicron zones also seem much more realistic (2.6\% at \rin\ and 15\% at \rout ). The emission in the northern and southern parts of the disk are now much better matched by the polarimetric models, which is indicated by the improved \chisqR\ of 7 (15 in the 2-zone model). Since the total \chisq\ decreases by $\sim$3600 using only 3 more parameters, the 3-zone model is clearly preferred by the data. 

However, the residuals in the difference images suggest that model improvements could be made by adding more parameters. The residuals show emission from the back side of the disk rim beyond 50~au, in the $K_s$-band total intensity image. This could be compensated for by submicron grains with a slightly different composition so that they maintain good total flux and polarimetric matches. This is not possible with our grain models. The polarimetric residuals in the $H$-band show too little forward scattering (see the East and South sides of the disk). Again, this could be addressed by micron-sized grains, which don't invoke too much forward scattering in the $K_s$-band image. We find that individual fits to either the $K_s$-band image or the $H$-band are much better than a combined fit, and this is mainly because a good balance between forward and backward scattering, or between submicron and microns grains is hard to achieve. The ALMA image model is clearly inadequate, as it is  unable to reproduce the sharp edge to the bright ring at ~0.5$''$. The smooth brighter edge in the model is created by a modest flaring (1.3\% at \rin\ and 7.5\% at \rout ). The observed sharp ring could be made by an additional dust component, shaped by a different phenomenon. The residuals also give evidence for asymmetries, which could be accounted for by disk ellipticity or local density perturbations. In summary, the models show that the major features of the disk emission can be described rather simply. 

\begin{figure*}
   \centering
   \includegraphics[width=0.9\hsize]{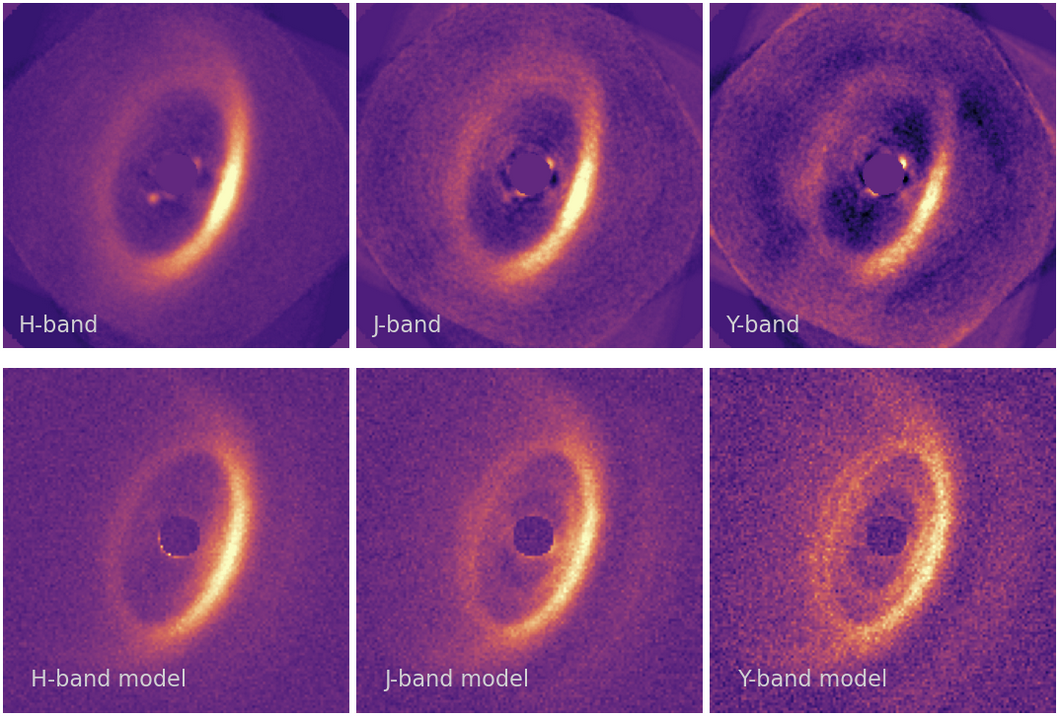}
      \caption{Comparison of Radmc3D 3-zone model images with $Y$,$J$ and $H$-band IFS observations, median combined over the typical band widths. For these, the models were not fitted because of computational limits. Simulated noise has been added to the data, but no planets were simulated. The data and model are morphological very similar. }
         \label{fig_comp_mod_yjh}
   \end{figure*}

In Figure~\ref{fig_comp_mod_yjh}, we compare the 3-zone model images with $Y$,$J$, and $H$-band IFS observations, median combined over the typical band widths. For these, the models were not fitted because of computational limits. However, the morphology and brightness of the model images are very similar to the data in all bands, suggesting reasonable matches. The $Y$-band match is poorest, showing "patchy" residuals, seemingly due to PSF halo mismatch at large scales. Notably, the dominance and degree of forward-scattering over backward scattering, the curvature at the outer part of the disk, and the depth and clarity of the disk gap are satisfactorily reproduced.

\subsection{Constraints on Model Parameters}

Calculating uncertainties is a challenge because we cannot fully rely on our results from Monte-Carlo style Bayesian estimation (using the Python $emcee$ package). This is because we were only able to run $\sim$38,000 samples (see Appendix~\ref{app_emcee}) given that each model takes 1 minute to create. This is just an unavoidable computing power limit, given the very high resolution of our data. In our estimate, the Bayesian uncertainties are too small. Thus, we attempted to estimate the uncertainty of our best-fit in the local minimum. To achieve this, we first computed the \chisq\ for each parameter by varying it along a linear grid, flanking the optimal fit value. During this process, we allowed the other 16 parameters to adjust accordingly, to identify a set of best-fit solutions. The uncertainty for each parameter was then determined based on the range within which the change in \chisq\ remained below 1\% of the minimum value. This 1\% threshold was chosen because the uncertainty in our \chisq\ estimates themselves are around 0.5\%. Although we cannot claim that we have found a true best-fit model, or robust uncertainties, we do show whether these constraints from the local minimum are weak or strong. 

\begin{table}[ht!]
\centering
\caption{3-zone Model best-fit Parameters}
\label{tab_mod2_prams}
\begin{tabular}{l r r c}
\hline\hline
Parameter & Value & Uncertainty & Units \\
\hline
\mdiska & $2.3e-3$ &  $+11\%,\ -17\%$ & \msun \\
\mdiskb  & $4.3e-4$ & $+2\%,\ -1\%$ & \msun \\
\mdiskc  & $4.3e-2$ & $+7\%,\ -16\%$ & \msun \\
\rin & $49.7$ & $+1\%,\ -3\%$ & au \\
\rout & $91.4$ & $+4.3\%,\ -0.1\%$ & au \\
$PA$ & $161.4$ & $+1\%,\ -2\%$ & deg \\
$i$ & $55$ & $+1\%,\ -1\%$ & deg \\
\plh & $2.9$ & $+4.9\%,\ -2.4\%$ & - \\
\plhc & $0.252$ & $+10\%,\ -24\%$  & - \\
\hr & $0.026$ & $+1\%,\ -7\%$ & - \\
\hrc & $0.013$ & $+7\%,\ -16\%$ & - \\
\plsig & -1.1 & $+37\%,\ -9\%$ & - \\
\plsigc & 4.88 & $+3\%,\ -23\%$ & - \\
$a1$ & 0.1 & $+14\%,\ -1\%$ & \mic \\
$a2$ & 10.9 & $+13\%,\ -5\%$ & \mic \\
$a3$ & 300 & $+1000\%,\ -17\%$ & \mic \\
Gap\ clearing & 0.0143 & $+4\%,\ -32\%$ & \mic \\
Dust1 & Olivine & - & - \\
Dust3 & Pyroxene & - & - \\
\hline
\end{tabular}
\end{table}

\begin{table}[ht!]
\centering
\caption{Goodness-of-Fit Statistics for the 3-zone disk model. Here $N_{dp}$ indicates the equivalent number of data points, or the number of resolution elements that have flux with a SNR above 2 in the image. The \chisq values are reduced \chisq\ for flux and image parts calculated separately (see appendix~\ref{app_goodfit}).}
\label{tab_gof_stats02}
\begin{tabular}{l r | l r | l r | l r}
\hline\hline
\multicolumn{2}{c|}{$H$, Stokes Q} & \multicolumn{2}{c|}{$H$, Stokes U} &\multicolumn{2}{c|}{$K_s$-band} & \multicolumn{2}{c}{870\mic} \\
\hline
$N_{dp}$  & 251 & $N_{dp}$  & 272 & $N_{dp}$  & 448 & $N_{dp}$  & 1279 \\
$\chi^2_{img}$  & 7 & $\chi^2_{img}$ & 6 & $\chi^2_{img}$ & 14 & $\chi^2_{img}$ & 4.5 \\
$\chi^2_{flux}$ & 0.1 & $\chi^2_{flux}$ & 0.13 & $\chi^2_{flux}$ & 0.01 & $\chi^2_{flux}$ & 0.13 \\
\hline
\end{tabular}
\end{table}

The uncertainties of the 3-zone model are shown in Table~\ref{tab_mod2_prams}. The majority of the parameters are extremely tightly constrained at the local minimum, implying that we get very poor fits for slight variations (1--2\%). These strong constraints are possible because of the high SNR and resolution of our images. Some highly constrained parameters are \rin, \rout, $PA$, $i$, and the mass in Zone2 (the micron-sized grain disk). The strong Zone2 mass-constraint comes from the delicate forward and back-scattering balance of the NIR images. We also note relatively strong constraints ($\sim$5\%) on the scale height and flaring parameter $\beta$ for Zone1 and Zone2, which just means the flaring in the disk is real and well-detected. Flaring in Zone3 is only needed to make the outer edge of the disk brighter, but it is also constrained to be small. The density power-law index of Zone1 and Zone2  is strongly constrained to have a  minimum steepness, and is found to have the typical range for flared disks. There is a strong upper limit of the gap clearing fraction ($0.014^{4\%}_{-32\%}$), indicating the gap is indeed largely cleared of dust. The lower limit, although softer, also indicates that some dust is needed. Soft constraints on the Zone1 and Zone3 masses and the Zone3 grain size are likely due to optical thickness requirements being soft. Once a certain optical thickness is reached, the images do no change much. On the other hand, the disk must be optically thick in the NIR to mask its far side and to appear flared. Likewise, the Zone3 grains, which make a flat disk, must not influence the NIR images, and therefore must be sufficiently large (and cold). But given their large size, and low total surface area, we need a high total mass to match the intensity of the ALMA image.

\section{Discussion}

\subsection{Disk model} 
We have found a model with a reasonably good match to the dust disk around PDS 70, which we demonstrate by calculating the \chisq\ from our highest-SNR images. The most likely morphology is that of a highly flared disk with a deep gap within $\sim$ 50 au of the star. Moreover, the disk flaring is clearly detected in the $H$-band Stokes Q and U images. The NW outer disk arc seen  in $JHK$ observations is also naturally produced by the flared disk model (see \citeauthor{vanHolstein:2021}~\citeyear{vanHolstein:2021} and \citeauthor{Juillard:2022}~\citeyear{Juillard:2022} for other possibilities). Although we could not present a model that is completely consistent with the data in terms of a \chisq comparison, the nature of the residuals show that these are minor deficits of the model, and not fundamental limitations in our understanding of the physics. The most likely model limitations are  that we 1) varied the grain-size over a very coarse grid - 20\% per step, 2) used a single grain size for  each disk zone, not a size distribution, 3) did not try varying the ellipticity of the disk, and 4) used a monotonic dust density, except for the inner gap. In summary, although the search for a completely consistent fit is warranted, as this will help refine our description of the dust physics, there seems to be nothing mysterious about the model inadequacies. Thus the disk is flared in the NIR, flat in the \submm , and sub-micron to millimeter dust grains all occupy the same radial extent ($\sim$50 to 90~au). 

In the paper, we did not model the inner disk which was detected in \citet{keppler:2018}, and likely appears in our images as small features to the top-left and bottom-right of the coronagraph. These features are quite compact and irregular, and so would make the modeling complicated. We thus leave this effort to a later analysis.

We can compare the number of grains for each size to check whether their ratios agree with the grain size distribution expected from a collisional cascade $dN/da = a^\zeta$, with $\zeta=-3.5$ \citep{1969JGR....74.2531D}  (see~Appendix~\ref{app_grain_abund}). From the dust mass and grain size estimates of the 3-zone model (see Table\ref{tab_mod2_prams}), we find that the 10.9\mic\ grains are over-abundant by factor of $\sim$2 compared to the 0.1\mic\ grains, while the 300\mic\ grains are over-abundant by a factor of $\sim$1000. To reconcile the number ratios for Zone1 and Zone3, we need to set $\zeta=-2.67$. The sub-micron and micron grain numbers are both constrained by the NIR images. The total intensity images strongly constrain the micron-sized grains, demanding strong forward scattering and adequate optical thickness to give the disk a 3D appearance. The polarimetric images likewise demand both substantial forward and back-scattering. Furthermore, both grains populate the surface layers of the disk, thus indicate something about the true surface grain size distribution. These independent constraints make our $\zeta=-2.67$ estimate an important  measurement of the power-law for a proto-planetary disk, enabled by self-subtraction-free imaging with star-hopping.  

Taking the $\zeta=-3.5$ power-law as a reference, smaller grains are under-abundant, which suggests that their loss is efficient. Our shallower power-law may be typical of proto-planetary disks, as opposed to debris disks or the ISM \citep{1969JGR....74.2531D,draine_2006ApJ...636.1114D}. Although, \submm\ studies have yielded $\zeta \sim -3.5$ \citep{guidi_2022A&A...664A.137G}, comparing NIR and \submm\ image may tell a different story. Probable causes for the relative lack of small grains could be 1) small grain leakage into the disk gap \citep{2019ApJ...884L..41B}, 2) substantial grain-growth of sub-mm grains. Efficient grain growth in the disk mid-plane where the sub-mm grains settle could explain why the micron to sub-mm distribution is even shallower (see Figure~\ref{fig_grain_number_comps}). 

\subsection{Planets} 
We have found that planet $c$'s spectrum is very similar to planet $b$'s, while both spectra are much redder than that of the disk. This suggests that we are detecting the atmosphere, rather than a CPD for both planets. The CPDs, which are expected to be much colder ($<$ 500K, see \citeauthor{Wang:2021}~\citeyear{Wang:2021}) than the planet atmospheres, should have spectra more similar to the cold circumstellar disk's. As found by \citet{Wang:2021}, the evidence for the CPD comes from the emission at longer wavelengths ($M$-band and 855\mic\ detections). In the $JHK$-bands, planet $c$ is not majorly shrouded or contaminated by the main disk's dust. Thus planet $c$ is likely inside the gap as indicated by our orbital solution. It may seem that this fact should aid us in detecting planet $c$ in a polarized fraction map, but we showed this was not the case. This is primarily because the polarimetric image needs to be much higher SNR for this technique to work. 

To understand the planets' interaction with the gas and dust disk we look to the dynamical simulations by  \citet{2019ApJ...884L..41B}. They studied the mass accretion of the planet and their orbital evolution in a disk of gas and dust, tracking the locations of different particle sizes. Many of our findings can be explained by their dynamical model. They find that the planets will quickly settle into a 2:1 MMR in $\sim$ 0.1~Myrs, and can then remain stable for at least 2~$Myrs$. A 2:1 mean-motion resonance is consistent with our results, for example SMA$_b$ = 14 and SMA$_c$=22.2~au results in a 1:2 period ratio. These correspond to maximum separations from the star of 20.6 and 32.6~au ($=$SMA[1+e]), respectively. In both our models, we find that planet $c$ is clearly in the dust gap, and the apparent superposition with the disk is only a result of the inclination of the system. The \submm\ grains of the \citet{2019ApJ...884L..41B} simulations are confined near the gas pressure maximum, in a particularly narrow radial range when planet $c$ has a mass of 2.5\mjup\ which is closest to our estimate of 3.8$\pm$0.4\mjup. The location of this \submm\ ring is $\sim$ 75~au in the ALMA image, $63$~au in their simulation, but is dependent on the planet masses and initial semi-major axes (SMA =20~au and 35~au for planets $b$ and $c$ in their work). Presumably, an agreement could be found with the constraints in this work, if our SMA estimates were used. The smaller grains take a much more monotonic distribution 
than the \submm\ grains as we found in our best models. The dust density in their disk gap is also $\sim$1\% of that of the outer disk, while the small grains are leaking into the gap. 

In any case, we must ask why the disk flares sharply starting at $\sim$50~au, when the gas is spread vastly wider than the dust. The assumption is that the planets can keep the dust (especially the large grains) out of the gap, but not the gas. 
Both the  \citet{2019ApJ...884L..41B} simulations and the observations of \citet{Long:2018} and \citet{Facchini:2021} show that the gas density also drops significantly inside the gap. This should also decrease the disk scale height. Another way planet $c$ could do this is by producing extra gas turbulence near its orbit, especially if its orbit is eccentric.
A wider gap is cleared when the planets orbits have some eccentricity. However the eccentricity is only expected when the mass accretion onto the planets is $\sim$ to $7\times 10^{-7}$\mjup\  ~/yr, which is 10 times greater than the observed accretion rates \citep{Wagner:2018, Haffert:2019}. In summary, the observed morphology of the dust disk can be explained by the growth and orbital evolution of the planets.

\section{Conclusions}
In this work, we have presented $YJHK$--band imaging and spatially resolved spectroscopy, along with $H$-band polarimetry of the PDS 70 system, which is characterized by multiple planets and a young disk. Our observations have unveiled the true intensity of both the planets and the disk without the self-subtraction artefacts that are typical of ADI observations. This was achieved through the use of near-simultaneous reference star differential imaging, a technique also known as star-hopping. Our primary objectives were twofold: firstly, to explore the presence of new giant planets located beyond separations of 0.1$"$; and secondly, to closely examine the disk's morphology to gain insights into its interactions with the orbiting planets. We used radiative transfer modeling using Radmc3D in an attempt to match the near-infrared, polarimetric  and \submm\ imaging data, consistent with measurement errors. The spectra of the planets were also extracted and subsequently compared to that of the disk. Before presenting our most significant findings, we summarize some basic notes here:
\begin{itemize}
    \item We found a good model of the PDS 70 dust disk as demonstrated by \(\chi^2\) calculations from high-SNR images.
    \item Our data constitute $>$2000 independent high SNR measurements, warranting a detailed model of the system. 
    \item Our model fits are likely not optimal due to these limitations:
    \begin{itemize}
        \item Grain-sizes were only varied over quite a coarse grid (20\% per step).
        \item A single grain size was used for each disk zone.
        \item Disk had circular symmetry (no ellipticity).
        \item Disk had monotonic dust density, except for inner gap.
    \end{itemize}
    \item In the $JHK$-bands, we detect the atmospheres of the planets with little contribution from CPDs.
    \item We find that the dust disk is much more radially confined than the range of the gas disk detection.
\end{itemize}

And finally, we list below the more significant findings of our work:

\begin{itemize}
    \item The disk morphology is highly flared with a deep gap within \(\sim 50\) au.
    \item The disk flaring is clearly detected in $H$-band Stokes Q and U images and is consistent with the total intensity images (sans self-subtraction).
    \item The NW outer disk arc seen in $JHK$-band observations is explained as the outer lip of the flared disk.
    \item The disk is flared in the NIR, flat in the \submm . The dust grains span \(\sim50\) to \(90\) au.
    \item The grain size distribution index is estimated to be $\zeta$ = -2.67 for submicron to \submm\ (as opposed to the canonical $\zeta$ = -3.5; \citeauthor{1969JGR....74.2531D}~\citeyear{1969JGR....74.2531D}).
    \item Our best-fit model is thus under-abundant in smaller grains.
    \item The probable causes of this are: small grain leakage or efficient grain-growth to \submm\ sizes.
    \item Planet $c$'s spectrum is similar to planet $b$'s, but redder than that of the disk.
    \item Planet $c$ is well inside the gap (roughly 18~au interior to the gap edge).
    \item Orbit fits for both planets suggest eccentric orbits (eccentricity $>$0.2), when co-planarity and dynamical stability constraints are not imposed (see \citeauthor{Wang:2021}\citeyear{Wang:2021}). 
    \item Disk starts at a small scale height of 2.6\% at $\sim$ 50~au, increasing to 15\% at $\sim$90~au. 
    \item It is possible that the eccentricity of the planet orbits is associated with greater gas turbulence, which manifests itself as a dust disk with large scale height. 
\end{itemize}

This study demonstrates that detailed information can be gleaned on the morphology of the proto-planetary disk of the PDS 70 system, and its dust grain sizes and probable compositions through detailed radiative transfer modeling. Moreover, we were limited by computational power from obtaining maximum information from the data. Self-subtraction-free polarimetric and total intensity imaging, along with high resolution \submm\ imaging from ALMA have enabled these advances. Particularly helpful in this study was the relative spectroscopy between the planets and disk, which would also not be possible without true total-intensity imaging. Finally, with our new imaging techniques we can reliably detect the planet motion every year, obtaining ever-more stringent orbital constraints. We foresee that much more will be learned about the interactions between the planets and their maternal disk, even with current capabilities: with multi-wavelength ALMA imaging, deeper polarimetric imaging in the NIR, higher resolution spectra with ERIS/VLT (hopefully aided by star-hopping), dynamical models of disk and planet interactions, and even more detailed radiative transfer models.

\begin{acknowledgements}
This work has made use of the SPHERE Data Centre, jointly operated by OSUG/IPAG (Grenoble), PYTHEAS/LAM/CESAM (Mar- seille), OCA/Lagrange (Nice), Observatoire de Paris/LESIA (Paris), and Observatoire de Lyon. We would especially like to thank Julien Milli at the SPHERE data center in Grenoble for the distortion-corrected reductions used for the astrometric measurements. This project has received funding from the European Research Council (ERC) under the European Union's Horizon 2020 research and innovation programme (PROTOPLANETS, grant agreement No.~101002188).
\end{acknowledgements}

\appendix

\section{$Q_\phi$ and $U_\phi$ images}
\label{app_qu_phi}
\begin{figure}
   \centering
   \includegraphics[width=\hsize]{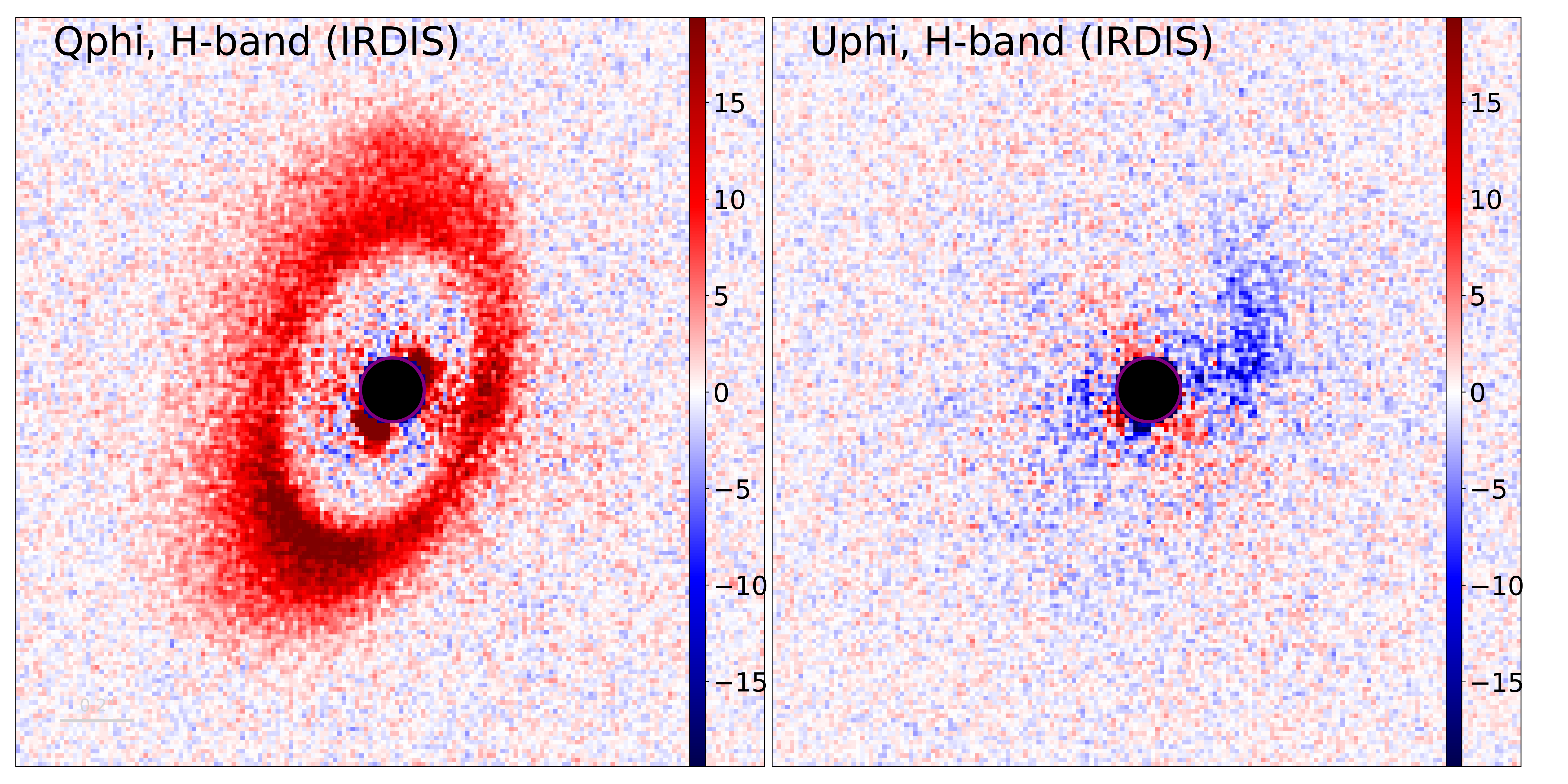}
      \caption{The $Q_\phi$ and $U_\phi$ images from the IRDAP pipeline \citep{vanHolstein:2020}.}
         \label{fig_qu_phi}
   \end{figure}

The $Q_\phi$ and $U_\phi$ images are also supplied by the IRDAP pipeline. They are made
from the Q and U polarimetric images according to the formulae given in \citet{deBoer:2020}.
They capture the intensity of the lightwaves with an electric field in the azimuthal direction. Although they should have a somewhat higher signal-to-noise ratio than the Q and U images (See \citeauthor{vanHolstein:2020}~\citeyear{vanHolstein:2020}), they make it more difficult to see the flaring of the disk in these images. We thus decided to model the Q and U images instead as they contain the same information and are more illustrative of the geometry of the disk. We show the $Q_\phi$ and $U_\phi$ images in Figure~\ref{fig_qu_phi} for purposes of completeness. 

\section{Goodness of fit}
\label{app_goodfit}
Here we describe how we estimated the goodness of model fits by calculating \chisq . For calculating probability distributions according to Bayesian Statistics, the usual total \chisq\ is used (see appendix~\ref{app_emcee}). 
For images, we calculate \chisq\ mainly for significantly detected parts of the image (see \citeauthor{Wahhaj:2021}~\citeyear{Wahhaj:2021}). Since the immediate background regions of any extended emission define its extent, these regions must be included in the \chisq\ calculation. 

Accordingly, we use the regions between 10 and 75 pixels or (125 $mas$ and 938 $mas$) from the star. We designate the area of this signal and background region as $A_{s+b}$. The noise per pixel is calculated with a typical robust sigma algorithm, and designated $\sigma$. The significant areas of detection are defined as all regions with signal twice this sigma value, which we call $A_s$. The area of a resolution element is designated $A_{resel}$, the diffraction limit at the observing wavelength or the beam size in \submm\ observations. Thus, our total \chisq\ for an image, re-scaled to reflect only the significant area is  
$$ \chi^2_{img} = \frac{A_s}{A_{resel}\ A_{s+b}} \sum_{i} \left(\frac{f_i - c.f'_i}{\sigma}\right)^2$$
where $i$ represents the pixels in $A_{s+b}$ and $f_i$ and $f'_i$ are the data and model pixel flux, respectively. Since the sum also yields an area in pixels, $\chi^2_{img}$ remains a unit-less fraction as intended. We designate $N_d = A_s/A_{resel}/ A_{s+b}$ as the number of effective data points.
The pixel size in the SPHERE images and our sampled ALMA images is the same.
Note that $c$ is a free scaling factor applied to the model to minimize the residual in the difference image, $data-c\times model$, over the $A_{s+b}$ region. Thus, the total image flux is ignored in this term; only the relative pixel-to-pixel fluxes affect $\chi^2_{img}$. 

This is because the total flux of the image has a much larger uncertainty of 15\%, which is typical of relative flux calibration in comparison to the star. Thus, we calculate the total flux uncertainty separately as 
 $\chi^2_{flux} =   ((f - f')/0.15)^2$, where $f$ and $f'$ are the total fluxes of the data and model, respectively. Hence, the total \chisq\ for all our data sets which are used for probability calculations is 
$$ \chi^2_{tot} =\sum_{j}{\chi^2_{j,\ img} + \chi^2_{j,\ flux} } $$
where $j$ represents the data sets ALMA 870\mic , SPHERE $K_s$-band, SPHERE Stokes Q in $H$-band, etc. For best-fit searches, we use the mean reduced \chisq\ or 
$$\frac{1}{N}\sum_{j}\frac{\chi^2_j}{{N_{d, j}}}$$ 
where $N=\sum_{j}2j$ (each dataset $j$ has 2 parts, image and flux), since this gives better interim fits before a search has converged, as explained in the main text.

The best 2-zone model parameters (Table ~\ref{tab_mod1_prams}) and \chisq\ values (Table ~\ref{tab_gof_stats01}) are presented here for completeness. The relevant discussion can be found in section~\ref{sec_mod_fit_res}.

\begin{table}[ht!]
\centering
\caption{2-zone Model best-fit Parameters}
\label{tab_mod1_prams}
\begin{tabular}{l r c}
\hline\hline
Parameter & Value & Units \\
\hline
\mdiska & 1.75$\times 10^{-5}$ & \msun \\
\mdiskb  & 0.024 & \msun \\
\rin & 51.5 & au \\
\rout & 86.7 & au \\
$PA$ & 160.3 & deg \\
$i$ & 51.4 & deg \\
\plha & 4.7 & - \\
\plhb & 3.4 & - \\
\hra & 0.068 & - \\
\hrb & 0.013 & - \\
\plsiga & 1.16 & - \\
\plsigb & 5.02 & - \\
$a1$ & 1.1 & \mic \\
$a2$ & 2200 & \mic \\
Dust & Olivine & - \\
\hline
\end{tabular}
\end{table}

\begin{table}[ht!]
\label{tab_gof_stats01}
\centering
\caption{Goodness-of-Fit Statistics for the 2-zone disk model. Here $N_{dp}$ indicates the equivalent number of data points, or the number of resolution elements that have flux with a SNR above 2 in the image. The \chisqR values are reduced \chisq .}
\begin{tabular}{l r | l r | l r}
\hline\hline
\multicolumn{2}{c|}{$H$-band} & \multicolumn{2}{c|}{$K_s$-band} & \multicolumn{2}{c}{870\mic} \\
\hline
$N_{dp}$  & 251 & $N_{dp}$  & 448 & $N_{dp}$  & 1279 \\
Image \chisq & 15 & Image \chisq & 14 & Image \chisq & 3.9 \\
Flux \chisq & 0.03 & Flux \chisq & 0.24 & Flux \chisq & 1.8 \\
\hline
\end{tabular}
\end{table}

\section{Radmc3d Models Details}
\label{app_radmc}
The radmc3d models are given the following resolutions in their spherical coordinate system (with coordinates, $r, \phi, \theta$). The radial direction is divided into two segments extending 2 to 30~au and from 30 to 200~au. Each segment is given a model resolution of 20 elements. The $\theta$ coordinate is segmented with bounds: 0, $\pi$/3, $\pi$/2 to 2$\pi$/3, and $\pi$ , with resolutions of 10, 20, 20, and 10 for the respective segments. The denser regions of the model are given higher resolutions. The $\phi$ coordinate is given a uniform resolution of 90 elements. We use 100,000 photons and 1,000,000 scattering events. Once a good fit is found, the model is run at a higher resolution to re-check the quality of fit. The higher resolutions are 20,40 elements for $r$,   20,40,40 and 20 elements for $\theta$, and 180 elements for $\phi$, reported in the same order as before. This time we use 300,000 photons and 1,000,000 scattering events.

\section{Constraining model parameters using $emcee$} 
\label{app_emcee}
Here we describe our attempt to constrain the model parameters by using the Bayesian estimation python package   {\it emcee}. 
The $emcee$ package samples the space of model parameters (17 in the case of the 3-zone disk), according to the probability of the model $e^{{-\chi^2}/2}$ using the Affine Invariant Ensemble Sampler \citet{Foreman-Mackey:2013}. The total \chisq\ is used here, as opposed to reduced \chisq . This is important to note, as the relative model probability and thus the strength of the constraints is strongly influenced by the number of independent measurements, and thus the magnitude of  \chisq. 

We used 35 parallel random walkers, each of which took 1090 steps, for a total of 38150 steps . For each walker, 242 initial steps were discounted as {\it burn-in}. The mean coherence time for walkers was $\sim$100 steps, but this was gradually increasing. This suggests that the probability space may be too complex, even for the adaptive {\it EnsembleSampler} routine we employed. Each model took an average of 56 seconds to create, and the sampler was run for 26 days to produce the constraints presented below. 

\begin{figure*}
\centering
\includegraphics[width=0.9\textwidth]{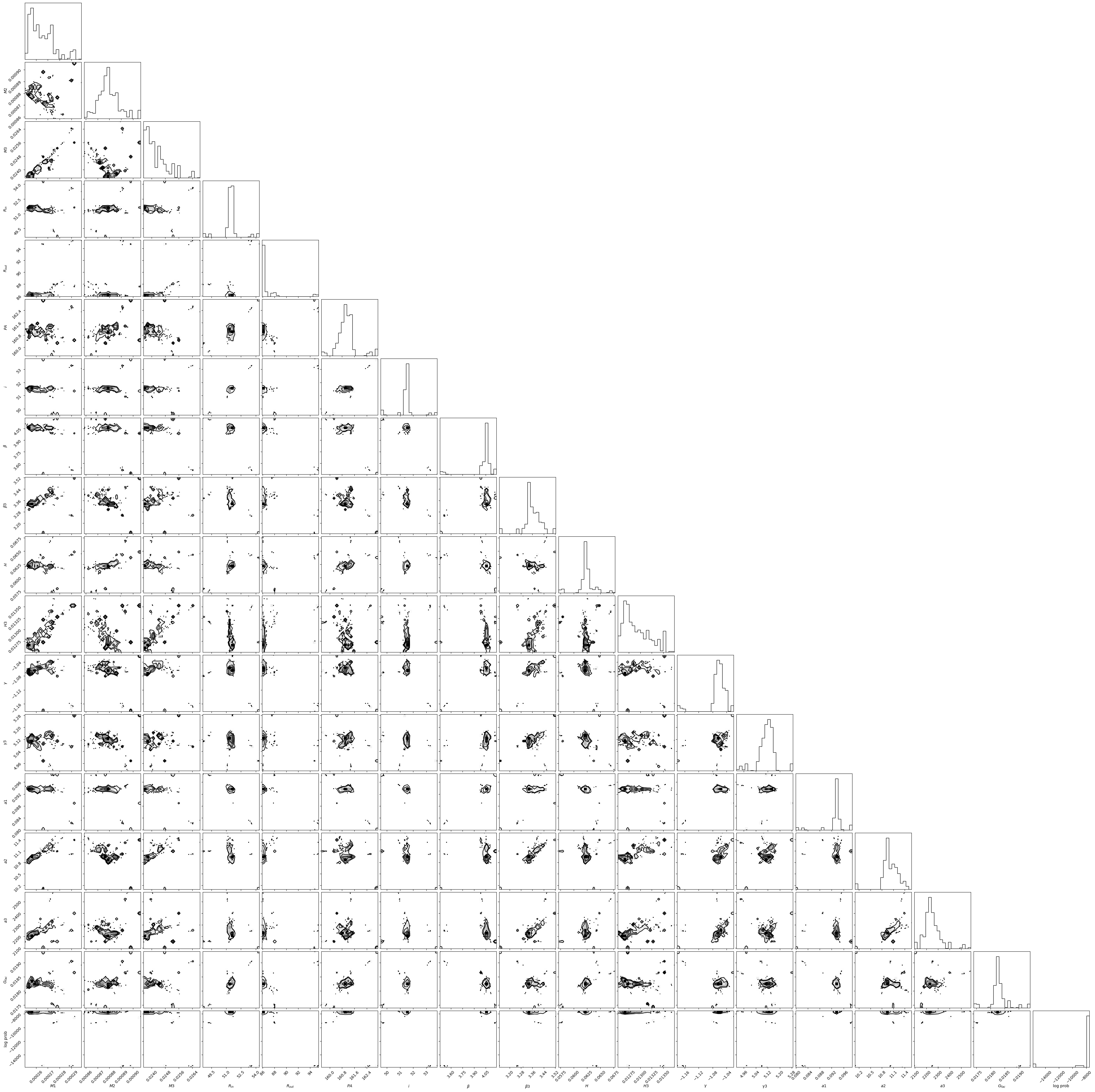}
\caption{Model parameter probability distributions from $emcee$. Convergence not reached after 26 days.}
\label{fig_emcee_corner}
\end{figure*}

Figure~\ref{fig_emcee_walkers}, shows the parameter values sampled for all the walkers. General convergence is noted for all parameters, but multimodal distributions could be indicated for some. In Figure~\ref{fig_emcee_corner}, we show the probability distributions for all parameters, and the 2D probability maps for all possible pairs of the 17 parameters. Seemingly all parameters are well-constrained (see Table~\ref{tab_emcee_constraints}). However, we know this to be unlikely, as some parameters are really only loosely constrained (see the large fractional error bars in Table~\ref{tab_mod2_prams}). For example, the upper limit to the size of the \submm\ grains are not constrained by the data as explained in section~\ref{sec_mod_fit_res}. In general, the constraints found seem artificially small when compared to manual exploration of \chisq values around the local minima. Again, this is because the walkers did not complete enough steps. 

\begin{table}[ht!]
\centering
\caption{Model parameter constraints from \textit{emcee} seem artificially small. This suggests that the search did not complete enough steps.}
\label{tab_emcee_constraints}
\begin{tabular}{l r r c}
\hline\hline
Parameter & Value & Uncertainty & Units \\
\hline
\mdiska  & 0.00026 & $+4.3\%,\ -2.7\%$ & \msun \\
\mdiskb  & 0.00088 & $+0.84\%,\ -0.8\%$ & \msun \\
\mdiskc  & 0.024 & $+3.1\%,\ -1.7\%$ & \msun \\
\rin & 51 & $+0.24\%,\ -0.39\%$ & au \\
\rout & 86 & $+1.8\%,\ -0.17\%$ & au \\
PA & 161 & $+0.19\%,\ -0.24\%$ & deg \\
$i$ & 51 & $+0.21\%,\ -0.3\%$ & deg \\
\plh & 4.1 & $+0.66\%,\ -0.71\%$ & - \\
\plhc & 3.4 & $+1.6\%,\ -0.64\%$ & - \\
\hr & 0.062 & $+1.2\%,\ -0.78\%$ & - \\
\hrc & 0.013 & $+2.9\%,\ -1.5\%$ & - \\
\plsig & -1.1 & $+1.4\%,\ -1.1\%$ & - \\
\plsigc & 5.1 & $+0.69\%,\ -0.76\%$ & - \\
a1 & 0.094 & $+0.54\%,\ -0.75\%$ & \mic \\
a2 & 11 & $+2.1\%,\ -1.1\%$ & \mic \\
a3 & 2200 & $+3\%,\ -1.7\%$ & \mic \\
Gap clearing & 0.018 & $+0.87\%,\ -0.56\%$ & - \\
\hline
\end{tabular}
\end{table}

\begin{figure*}
\centering
\includegraphics[width=0.9\textwidth]{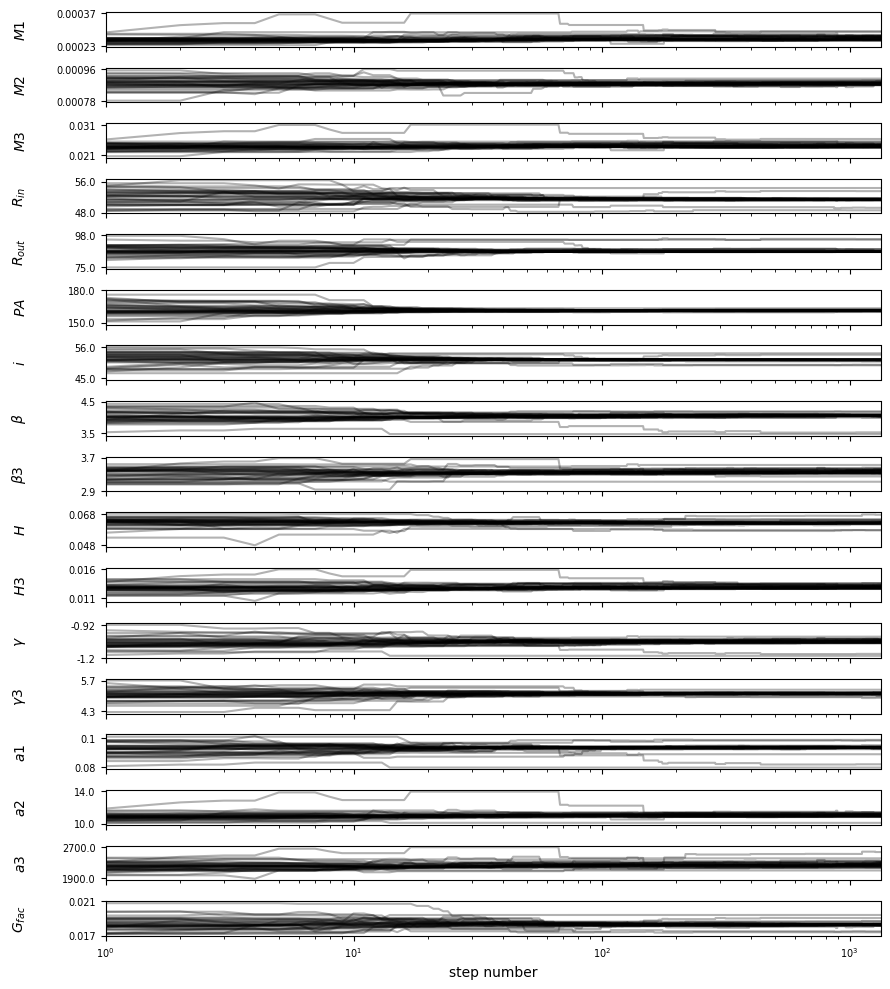}
\caption{The parameter values traced by the $emcee$ walkers. Convergence not reached after 26 days.}
\label{fig_emcee_walkers}
\end{figure*}

\clearpage  

\clearpage  

\section{Orbit fitting} 
\label{app_orbit_fitting}

In this section, we present the probability distributions of orbit fitting parameters discussed in section~\ref{sec_orbit_fitting}.

 \begin{figure}
  \centering
  \includegraphics[width=\hsize]{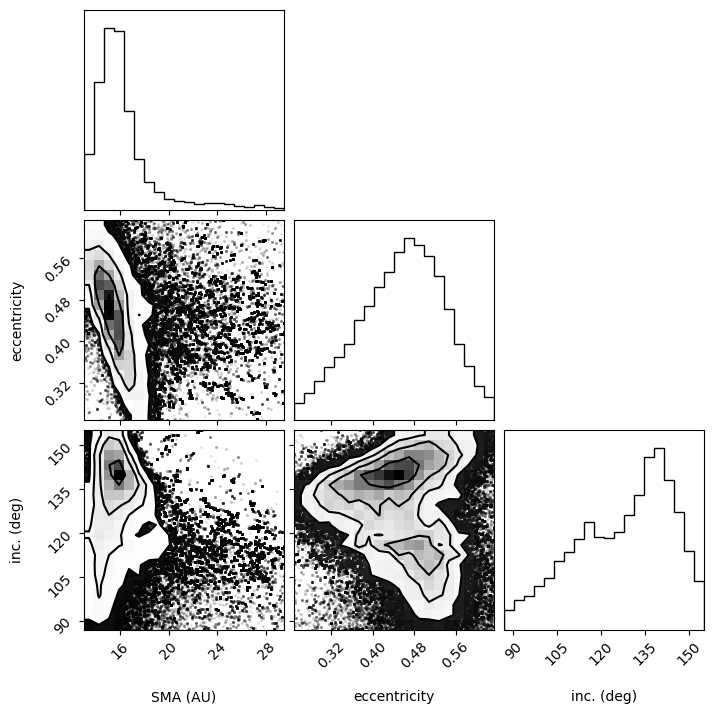}
     \caption{Probability distributions for the main orbital parameters of planet $b$, showing the 95\% confidence interval (see Figure~\ref{fig_orbit_planet_b_and_c}). It seems that a circular orbit is unlikely. The inclination may still be co-planar with the disk ($i \sim 49.3^o or\ 180^o-49.3^o = 130.7^o$).}
        \label{fig_orbit_planet_b_corner}
  \end{figure}

 \begin{figure}
  \centering
  \includegraphics[width=\hsize]{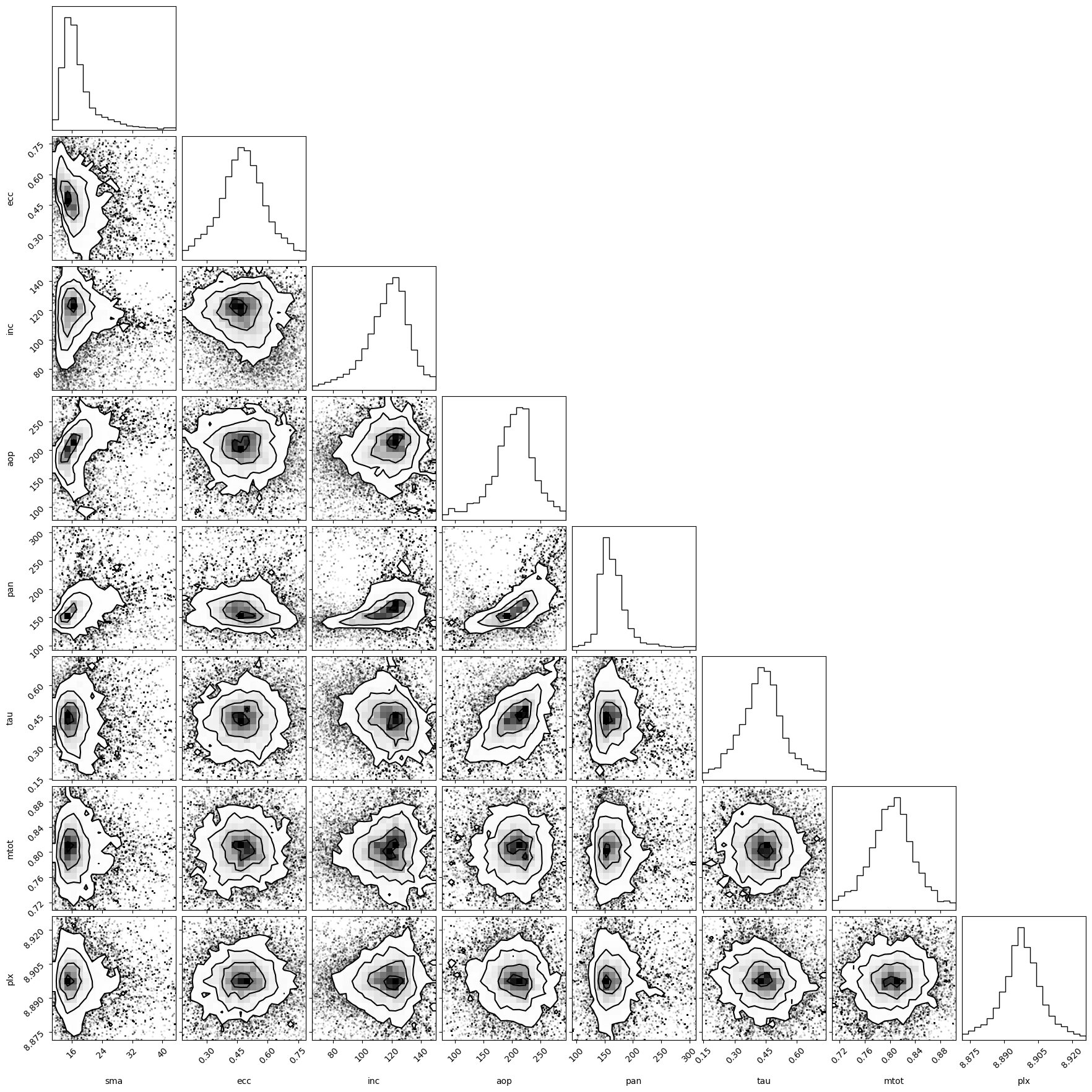}
     \caption{Probability distributions for all orbital parameters of planet $b$, showing the 95\% confidence intervals. The parameters are $sma$ or semi-major axis, $ecc$ or eccentricity, $inc$ or inclination, $aop$ or argument of periastron, $pan$ or position angle of nodes, $tau$ or epoch of periastron passage (expressed as a fraction of orbital period past a specified offset), $mtot$ or system mass, and $plx$ the system parallax.}
        \label{fig_all_orbit_planet_b_corner}
  \end{figure}

 \begin{figure}
  \centering
  \includegraphics[width=\hsize]{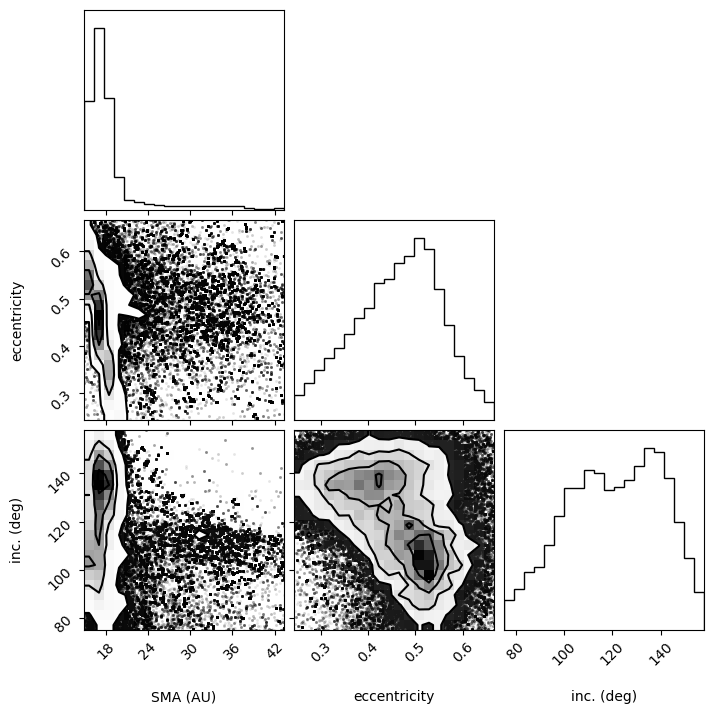}
     \caption{Probability distributions for the main orbital parameters of planet $c$, showing the 95\% confidence interval (see Figure~\ref{fig_orbit_planet_b_and_c}). It seems that a circular orbit is unlikely. The inclination may still be co-planar with the disk ($i \sim 58.5^o or\ 180^o-58.5^o = 121.5^o$).}
        \label{fig_orbits_c_corner_plot}
  \end{figure}

 \begin{figure}
  \centering
  \includegraphics[width=\hsize]{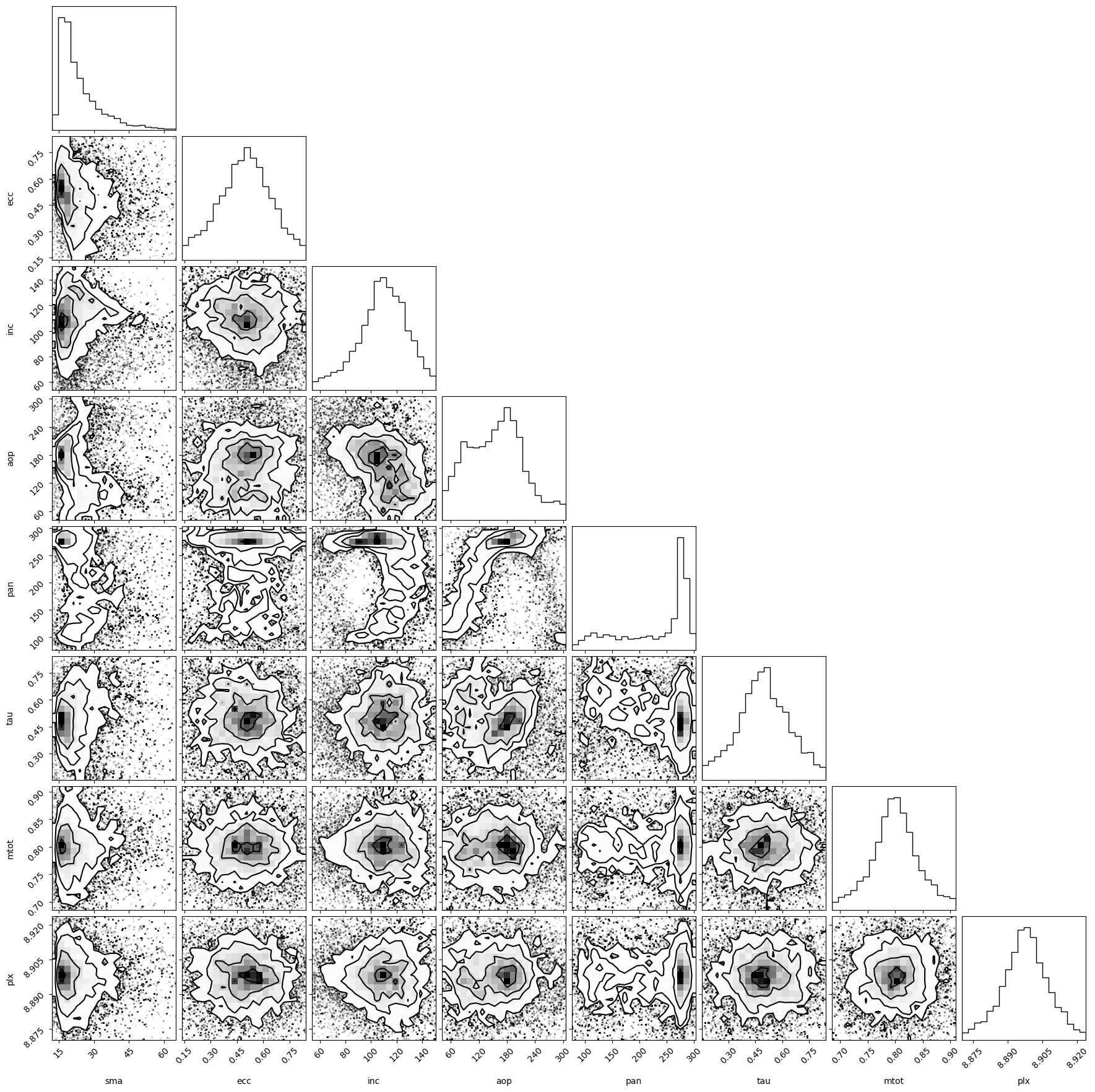}
     \caption{Probability distributions for all orbital parameters of planet $c$, showing the 95\% confidence intervals. The parameters are $sma$ or semi-major axis, $ecc$ or eccentricity, $inc$ or inclination, $aop$ or argument of periastron, $pan$ or position angle of nodes, $tau$ or epoch of periastron passage (expressed as a fraction of orbital period past a specified offset), $mtot$ or system mass, and $plx$ the system parallax.}
        \label{fig_all_orbit_planet_c_corner}
  \end{figure}

\section{Grain abundance ratios} 
\label{app_grain_abund}

In this section, we compare the grain abundance ratios between the three grain populations used in this paper. Let us consider an infinitesimal range of grain sizes, $\Delta a$, within which to count the grains. According to a single power-law size distribution (e.g. $dN/da = n \, a^{-3.5}$;\ \citeauthor{1969JGR....74.2531D}~\citeyear{1969JGR....74.2531D}), the number of grains within this range is 
\[\Delta N_i=n \, a_i^{\eta} \, \Delta a.\]
Note that $\Delta a$ is an absolute bin size such that $\Delta a \ll a$, but it is independant of $a$. \footnote{This is a computational requirement set by the power-law $dN/da = n \, a^{\gamma}$, and not related to whether the models demand values proportionally or absolutely close to $a$.}

Thus, the number ratio of the grainsizes $a_i$ and $a_j$ within a bin size $\Delta a$ is given by
\begin{equation}
\Delta N_i/\Delta N_j= \left(\frac{a_i}{a_j}\right)^{\eta}
\label{eqn_grain_number_ratio}
\end{equation}
Now we can compare these number ratios to the model-estimated total masses for each dust population, $M_i$ (see Table~\ref{tab_mod2_prams}): 
\[M_i=\rho \, V_i \, \Delta N_i \]
\[\Rightarrow \Delta N_i = \frac{M_i}{\rho \, V_i} \; \propto M_i \, a_i^{-3}\]

Here, $V_i$ is the grain volume, $\frac{4}{3}\, \pi \, (a_i/2)^3$, and $\rho$ is the grain density. Thus, the ratio of the grain sizes from the model-mass estimates is 
\[\Delta N_i/\Delta N_j = \frac{M_i}{M_j} \left(\frac{a_i}{a_j}\right)^{-3}.\]
Using this result in Equation~\ref{eqn_grain_number_ratio}, we solve for $\eta$:
\[\eta = \frac{log(M_i/M_j)}{log(a_i/a_j)}-3.\]

\begin{figure}
  \centering
  \includegraphics[width=\hsize]{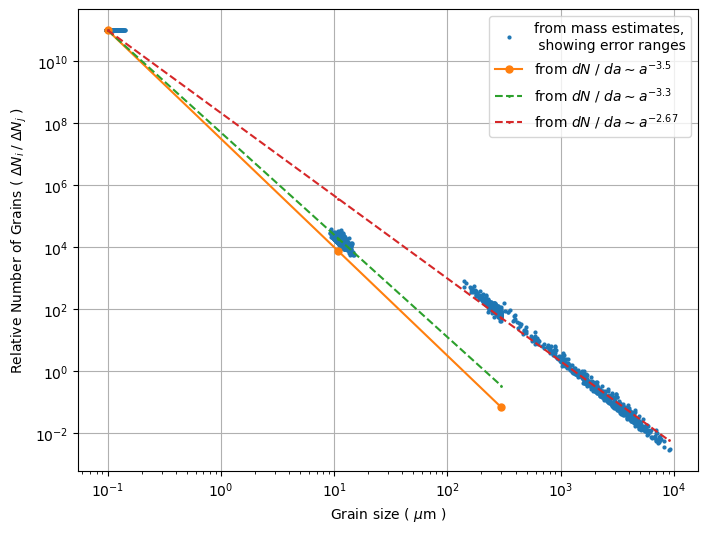}
     \caption{The model-estimated relative abundances of the three grain sizes compared to those predicted by the power-law $dN/da \propto a^{-3.5}$ \citep{1969JGR....74.2531D}. We used the asymmetric upper and lower uncertainties (see Table~\ref{tab_mod2_prams}) to generate 3000 data pairs $a_i$ and $\Delta N_i$, in order to properly account for correlated errors.}
        \label{fig_grain_number_comps}
\end{figure}

Using the estimates with upper and lower error bars in Table~\ref{tab_mod2_prams} and comparing number ratios between $0.1$ and $10.9$\mic\ grains we get $\eta=$ -3.33 to -3.27, while comparing $0.1$ and $300$\mic\ grains we get $\eta=$ -2.69 to -2.66. Using the equations above, we see that 10.9\mic\ grains are over-abundant by a factor of 2, compared to the 0.1\mic grains (assuming $dN/da \propto a^{-3.5}$), while the $\sim$300\mic grains are over-abundant by more than a factor of 1000. In Figure~\ref{fig_grain_number_comps}, we show that the discrepancies with the $dN/da \propto a^{-3.5}$ power-law are quite significant for the $\sim$300\mic\ grains.

\bibliographystyle{aa}

\end{document}